# Distributional welfare impacts and compensatory transit strategies under NYC congestion pricing

Xiyuan Ren*, Zhenglei Ji, Joseph Y. J. Chow
*C2SMART Center, New York University Tandon School of Engineering, Brooklyn, USA*
* Corresponding author: xr2006@nyu.edu

**Acknowledgments:** Funding support from C2SMARTER is appreciated. Data shared by Replica Inc. are gratefully acknowledged.

**Declaration of interest**: The authors declare that there are no competing interests related to this work.

## ABSTRACT

Early evaluations of NYC's congestion pricing program indicate overall improvements in vehicle speed and transit ridership. However, its distributional impacts remain understudied, as does the design of compensatory transit strategies needed to mitigate potential welfare losses. This study identifies population segments and regions most affected by congestion pricing, and evaluates how those welfare losses can be compensated through transit improvements funded by the toll revenues. We estimate joint mode and destination models using aggregated synthetic trips in the NY–NJ–CT–PA Combined Statistical Area (CSA) and calibrate toll-related parameters using post-toll changes reported by MTA. Welfare impacts of congestion tolls are measured as changes in consumer surplus (CS) before and after program implementation. Compensatory transit strategies are evaluated by quantifying the reductions in transit wait time and fare discounts required to offset the CS losses. The results show that the program leads to an accessibility-related CS loss of $397.23 million per year, while generating net passenger toll revenue of $523.44 million per year estimated based on the MTA's report—indicating a net welfare gain. However, these gains in benefits conceal significant disparities. CS losses are concentrated in upper Manhattan, Brooklyn, Queens, and Hudson County, particularly among travelers less able to shift to transit or alternative destinations. Achieving a general compensation requires modest investment—a 0.63-minute (13%) reduction in wait time or $165.15 million in annual fare subsidies for NYC residents, and a 2.12-minute (28%) reduction or $171.42 million for New Jersey residents. However, ensuring that no population group and county unit is made worse off is substantially more costly and infeasible through transit improvements alone. These findings underscore the need for differentiated compensation strategies: uniform fare discounts lead to overcompensation for some groups, whereas segment-specific discounts, origin-based fare reductions, or commuter pass bundles can achieve equitable accessibility restoration at lower fiscal cost.

## 1. Introduction

On January 5, 2025, New York City Metropolitan Transportation Authority (MTA) launched the Central Business District Tolling Program (CBDTP), which is the first cordon-based congestion pricing scheme in the United States (Cook et al., 2025; National Bureau of Economic Research, 2025; Nogueira, 2025). The program charges vehicles entering the Congestion Relief Zone (CRZ), with tolls varying by time of day, vehicle type, and payment method (Metropolitan Transportation Authority, 2025a). Although its implementation follows decades of political debate and failed attempts dating back to proposals in the 1970s and Mayor Bloomberg's PlaNYC initiative in 2007 (Bloomberg, 2007; Schaller, 2010; Schwartz et al., 2008), the program now serves as a critical case study for assessing the economic, behavioral, and environmental impacts of congestion pricing in the U.S. context.

The first evaluation report published by Metropolitan Transportation Authority (MTA) points to positive benefits: vehicle entries into the CRZ decreased by 11%, while vehicle miles traveled decreased by 7.1%. Total greenhouse gas emissions decreased 6.1% year-over-year in the CRZ. In addition, the program generated $468 million in net revenue between January and October and is projected to exceed $500 million in its first year (Metropolitan Transportation Authority, 2026a). However, these gains have not quelled public opposition: at least ten lawsuits have been filed against the MTA and state officials by business coalitions, elected officials from New Jersey, and other stakeholders (Harris & Ley, 2024). Resistance is reinforced by longstanding narratives of necessity and fairness, as drivers question the feasibility of shifting away from car travel while lower-income groups and New Jersey commuters underscore the disproportionate financial burdens they face (Baghestani et al., 2022; Chen, 2025; Schaller, 2010). These debates highlight the tension between the program's demonstrated benefits and ongoing concerns about its potential negative effects.

Although congestion pricing should theoretically create net welfare gains when the revenues are redistributed back to users, redistribution mechanisms are unclear and outcomes may take years to observe. Consequently, measuring the distributional impacts of reinvesting toll revenue in transit systems becomes essential for achieving both efficiency and equity goals (Basso & Jara-Díaz, 2012; Chen & Nozick, 2016; Marazi et al., 2024). Such measures not only ensure equitable opportunities for disadvantaged groups but also encourage a broader mode shift that yields environmental benefits and long-term public trust (Isaksen & Johansen, 2025). Achieving these outcomes, however, requires rigorous quantitative evaluation of how congestion pricing influences travel behavior and how revenues can be redistributed to compensate affected populations.

While welfare analysis based on logsum utilities from discrete choice models (DCMs) has been widely applied in transportation research (He et al., 2021; Ji, 2025; Li et al., 2021), the absence of consistent trip data for the broader metropolitan area, as well as the lack of models that capture shifts in traveler preferences before and after the program implementation, remain key barriers to assessing compensatory transit policies under NYC's congestion pricing.

This study addresses the current research gap by combining synthetic data from Replica Inc. and observed data from MTA with advanced choice modeling techniques. The study area covers the NY–NJ–CT–PA Combined Statistical Area (CSA) (Wikipedia contributors, n.d.), spanning portions of four states to capture cross-state trip patterns and potential spillover effects from congestion pricing. To consider travelers' taste heterogeneity, we consider 16 segments defined by four mutually exclusive population groups (NotLowIncome, LowIncome, Senior, and Student), two trip purposes (Commute and Non-commute), and two time periods (Peak and Overnight[1]). Truck trips are excluded from the analysis, as truck drivers have distinct travel preferences and are less likely to benefit from public transit improvements.

We formulate market-level joint travel mode and destination choice models for both ex-toll and post-toll periods. For the ex-toll model (2023 Q2), we use synthetic population data provided by Replica Inc., comprising over 70 million trips in our study area on a typical weekday in the second quarter of 2023. Trips originating from the same county and made by the same segment are aggregated into markets, with each mode–destination pair constituting an alternative. Market shares are then calculated for each alternative. We estimate an inverse product differentiation logit (IPDL) model (Fosgerau et al., 2024), setting multinomial logit (MNL) (McFadden, 1972) and nested logit (NL) (McFadden, 1977) as benchmark models.

For the post-toll model (2025 Q2), we retain most settings from the ex-toll model while introducing toll-related parameters to capture the effects of congestion pricing on travel preferences. Traffic counts, for-hire vehicle trips, and transit ridership data from the MTA (2026a) are used to calibrate these parameters. The Moore-Penrose pseudoinverse (Moore, 1920) is applied to obtain the minimum-norm solution.

We then apply the models to quantify changes in trip welfare attributable to the congestion toll. Following existing mode and destination choice studies, trip welfare is proxied using consumer surplus (CS) calculated as the logsum of alternative utilities (Small & Rosen, 1981) and converted to monetary terms by dividing by the estimated travel cost parameter (Vij & Walker, 2016). Shapley value decomposition (Shapley, 1953) is employed to isolate the impacts of non-toll-related changes on trip welfare.

Finally, we pose the question: *where would revenues need to be distributed* in the transit system to make improvements to sufficiently compensate for the CS losses associated with the toll. Specifically, we calculate the increase in transit service frequency and the level of fare discounts necessary to compensate for the loss in CS. Moreover, we consider two perspectives: Kaldor–Hicks efficiency (Kaldor, 1939) and Pareto improving (Varian, 1992). The former seeks to compensate the aggregate CS loss, while the latter requires that no individual or traveler group is made worse off. Although our analysis does not measure specific operational cost for reducing transit wait time, it quantifies the CS compensation achieved at each level of wait time reduction, providing transit agencies with a roadmap for investment prioritization. To facilitate future research, we upload the codes, shareable datasets, and model results to a GitHub repository.

The remainder of the paper is organized as follows. Section 2 reviews studies on congestion pricing and welfare analysis based on DCMs. Section 3 outlines the methodology, including data processing, model estimation, and welfare measurement. Section 4 presents the results of joint mode and destination choice models, impact analysis, and strategy evaluation. Section 5 discusses the policy implications of our findings and limitations of our analysis. Section 6 concludes with key takeaways and directions for future research.

[1] According to the CBDTP pricing periods, the “Peak” period is defined as 5 a.m.–9 p.m., and the “Overnight” period covers the remaining hours.

# 2. Literature review

## 2.1 Congestion pricing: implementations, effectiveness, and concerns

Congestion pricing is designed to internalize the external cost of driving during peak periods, particularly the additional travel times imposed on others (De Palma & Lindsey, 2011; Downs, 2005). Landmark programs in London (Santos & Bhakar, 2006), Stockholm (Eliasson, 2009), and Singapore (Kockelman & Kalmanje, 2005) have demonstrated that charging drivers for access to constrained urban road space reduces traffic volumes, brings environmental benefits, and affects travel preferences. For instance, London's "Congestion Charge" policy reduced car miles by 20–25% during the charging period and reduced total vehicle miles traveled in the zone by 10–15% (Leape, 2006; Santos & Bhakar, 2006). Congestion pricing in Stockholm reduced $NO_x$ by about 8% and $PM_{10}$ by around 13%, while children's asthma attacks dropped significantly (Simeonova et al., 2021). A study in Singapore revealed that even a small toll can significantly decrease peak-time demand by shifting traffic to public transit and less congested hours (Wongpiromsarn et al., 2012), indicating that drivers are more likely to adjust either their departure time or travel mode when congestion tolls are applied.

New York City implemented the first cordon-based congestion pricing program in the U.S. in January 2025, targeting one of the most congested areas in the world. Toll rates start at $9 for passenger cars and small commercial vehicles with E-ZPass, $4.50 for motorcycles, and $14.40–$21.60 for trucks and buses, with a 75% overnight discount and surcharges of up to 50% for non-E-ZPass users billed by mail. Taxis and ride-hailing vehicles pay per-trip fees of $0.75 and $1.50, respectively. The program exempts emergency vehicles, provides partial credits for trips entering via tolled bridges and tunnels, and offers a 50% discount to low-income drivers after their first ten trips each month.

Despite the promising outcomes reported in early evaluations (Cook et al., 2025; Metropolitan Transportation Authority, 2026a), equity concerns have remained central to the policy debate. A Siena College Research Institute (2024) poll showed that statewide opposition rates were around 63% before its implementation. A key point of contention is whether the program disproportionately affects people from different regions and income groups. Critics argue that it may impose undue burdens on low-income drivers and residents of transit deserts who rely on driving and for-hire vehicles (Ghassabian et al., 2024). In contrast, Cook et al. (2025) find broadly similar speed improvements across income quintiles, with the largest gains occurring in lower-income areas. Beyond residents, truck drivers and freight companies have voiced strong concerns over the toll's financial burden, arguing that it threatens freight-dependent businesses (American Trucking Associations, 2025). Small business owners have also reported financial strains from both higher freight costs and declines in customer foot traffic (Shalma, 2025).

With all of these debates, the success of New York's congestion pricing depends on overcoming the dual challenge of demonstrating efficiency gains while ensuring that travelers perceive both the necessity and feasibility of shifting away from car use. This requires a deeper insight into how congestion tolls reshape travel behavior and influence overall trip welfare.

### 2.2 Welfare analysis based on DCMs

DCMs have been widely applied in transportation research to forecast travel demand by assuming travelers make decisions by maximizing the overall utility they can expect to gain (Bowman & Ben-Akiva, 2001). These models enable researchers to examine how attributes such as travel time, monetary cost, and convenience of transfer influence the probability of selecting specific modes and destinations (Hensher & Ho, 2016; Ren & Chow, 2022). Welfare impacts of transportation policies can be evaluated using consumer surplus (CS) derived from discrete choice models, where the logsum of utilities provides a measure of aggregate accessibility (Small & Rosen, 1981; Vij & Walker, 2016). For instance, Standen et al. (2019) explored the use of the logsum measure of CS for valuing the user benefits of new separated cycleways in Sydney. Ren et al. (2024) proposed a choice-based decision support tool for determining optimal service regions for on-demand mobility that balances revenue generation and welfare gains. CS in joint mode and destination models is often linked to “accessibility”, referring to the “ease” with which desired destinations may be reached (Niemeier, 1997). Bills et al. (2022) estimated a destination–mode (destination in the upper branch and mode in the lower branch) choice model to calculate logsum utilities and explored the equity impacts of a microtransit service in Metropolitan Detroit.

A growing body of literature has estimated DCMs and conducted welfare analysis in the context of congestion pricing. He et al. (2021) conducted a validated multi-agent simulation for NYC and demonstrated that congestion pricing produces heterogeneous demographic impacts, differing substantially across time periods and neighborhoods, highlighting the need for differentiated pricing strategies that account for varied commuter patterns and spatial traffic dynamics. Li et al. (2021) examined solutions for maximizing travelers’ welfare by varying toll levels and locations across road network in Austin, Texas. They suggested that congestion tolls can do more harm than good unless travelers shift out of the peak periods or revenues are returned to travelers as credits. Tarduno (2022) proposed a departure time choice model to evaluate second-best congestion pricing schemes in proposed cordon zones across several U.S. cities, showing that spatial leakage and imperfect pricing prevent the realization of welfare benefits. Ji (2025) estimated a mode choice model using NYC Citywide Mobility Survey data to examine the distributional impacts of NYC’s CBDTP, highlighting that raising tolls without reinvestment to public transit delivers negligible mode shift. Together, these studies underscore the importance of systematically examining distributional impacts and evaluating compensatory strategies to ensure that congestion pricing policies balance efficiency gains with equity considerations.

Despite numerous innovative ideas and valuable empirical findings, several critical research gaps remain in applying DCMs to evaluate NYC’s congestion pricing program. First, most existing studies rely on travel survey data from a single city, overlooking the broader regional impacts of New York City’s congestion toll, which extend to the greater metropolitan region across multiple states (Metropolitan Transportation Authority, 2026b). Second, relevant choice models either focus only on mode choice (Ji, 2025) or employ a nested structure by assuming unidimensional correlations (Bills et al., 2022), which may overlook the multidimensional substitution patterns across modes and destinations. Third, these studies typically treat travelers’ taste parameters as stable across pre- and post-implementation conditions, neglecting potential preference shifts induced by the program. Last but not least, limited attention has been given to policy scenario analysis, particularly regarding how congestion pricing revenues should be redistributed to travelers through compensatory strategies that offset the additional monetary burden, as well as the trade-offs under aggregate efficiency and distributional equity goals.

### 2.3 Our contributions

This study contributes to the literature on congestion pricing by addressing several of the key gaps identified above. First, we use synthetic trip data to incorporate a broader spatial scope including the NY–NJ–CT–PA CSA, thereby capturing the major regional spillover effects of the program. Second, we contribute to the joint mode and destination choice literature by introducing the IPDL model that extends the Berry–Levinsohn–Pakes (BLP) framework (S. Berry et al., 1995) to allow for nested and correlated utilities across alternatives within a market share modeling framework. Compared with other models in the generalized extreme value (GEV) family—such as nested logit (McFadden, 1977) and cross-nested logit (Vovsha, 1997)—IPDL relies on a linear regression representation of inverted market shares rather than full maximum likelihood estimation. This offers a more computationally efficient and scalable approach for estimating demand in settings with large choice sets. Third, we relax the common assumption of stable taste parameters by explicitly estimating traveler preferences before and after the implementation of congestion pricing, leveraging observed traffic counts to calibrate preference shifts induced by the toll.

In addition, this study advances policy evaluation by quantifying the role of compensatory transit strategies. We assess distributional impacts on accessibility through logsum utilities from the choice models disaggregated by demographic and trip segments. Building on this approach, we examine how increased transit service frequency and fare discounts can offset welfare losses attributable to congestion tolls. While empirical evaluation of such strategies would ideally rely on post-implementation data, the limited availability of such data—particularly in the early stages of a pricing program—necessitates a model-based approach. Our model framework offers a principled and internally consistent platform for conducting these counterfactual policy evaluations.

Collectively, these contributions position our study among the first to jointly assess the accessibility impacts and compensatory transit strategies of NYC's congestion pricing program within a joint mode–destination choice framework.

## 3. Data and methodology

### 3.1 Study area and data collection

#### *3.1.1 Study area*

While the toll cordon applies only to Manhattan south of 60th Street (also known as Congestion Relief Zone, CRZ), its impacts extend well beyond the city's boundaries (Metropolitan Transportation Authority, 2026b). Our study area therefore encompasses the broader NY–NJ–CT–PA Combined Statistical Area (CSA), spanning portions of New York, New Jersey, Connecticut, and Pennsylvania (Wikipedia contributors, 2026). By focusing on this broader metropolitan region (Fig. 1), we are able to account for cross-state commuting and potential spillover effects that may reshape the distributional impacts of congestion pricing.

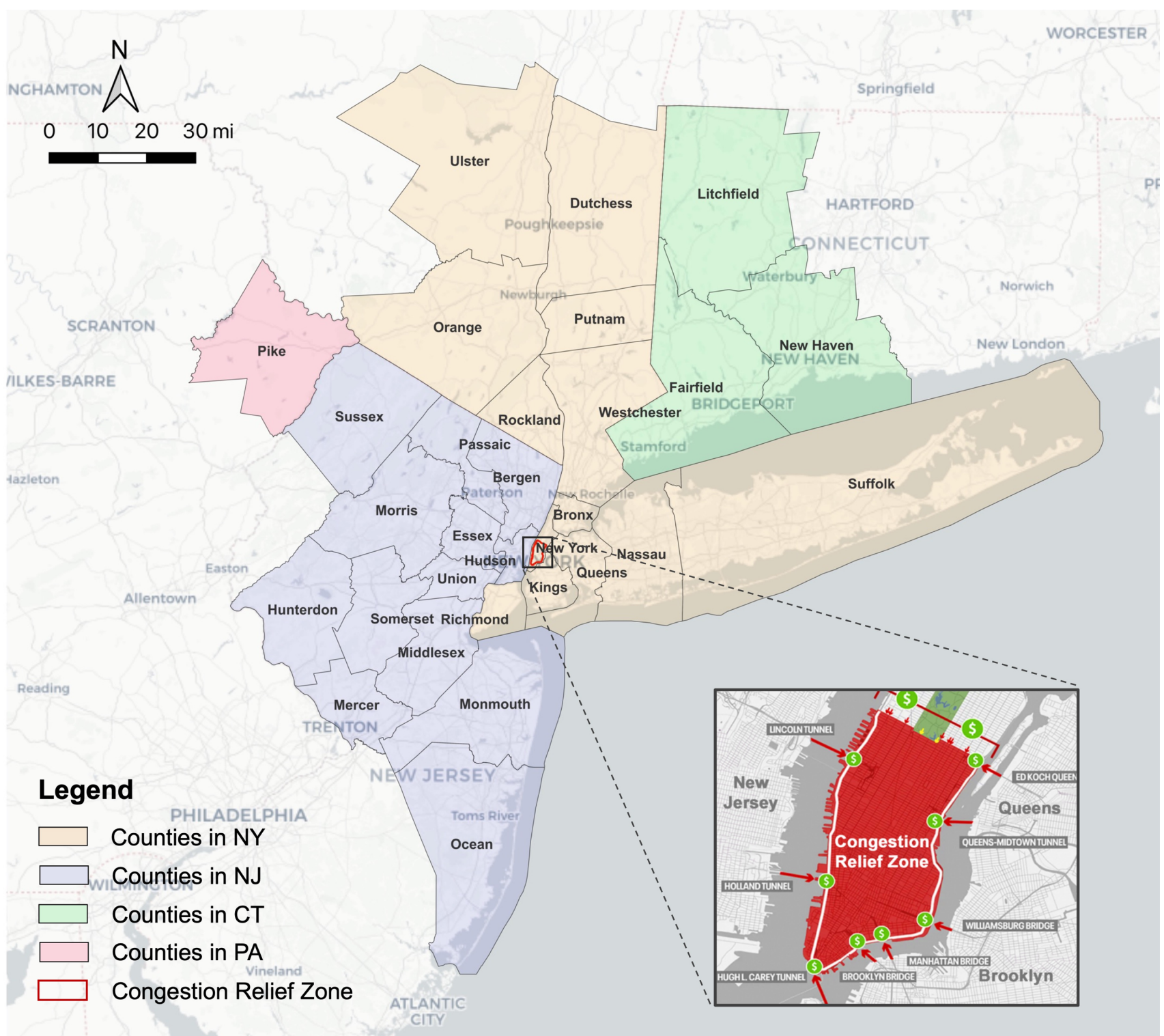

**Fig. 1.** Congestion Relief Zone (CRZ) and NY–NJ–CT–PA Combined Statistical Area (CSA). The inset map is adapted from a figure in Meier (2024).

*3.1.2 Synthetic trip data*

Consistent travel data remains scarce, particularly across different states and between urban and rural communities (Bachir et al., 2019; Parr et al., 2020). Synthetic population data help address this limitation by providing harmonized, large-scale trip records that capture travel behavior across regions in a consistent framework (Hörl & Balac, 2021).

We use synthetic population data provided by Replica Inc., containing over 70 million synthetic trips in our study area on a typical weekday in the second quarter of 2023. The dataset was generated through a combination of mobile phone data, census data, economic activity data, and built environment, representing large-scale travel behavior before the congestion pricing program (Replica Inc., 2024). For each individual trip, the dataset contains sociodemographic attributes, trip origin and destination at the census block group level, departure and arrival times, and travel mode—driving, public transit, for-hire vehicle (fhv), biking, walking, or carpool. To assess the reliability of Replica data, we compare inbound travel patterns to the Manhattan CBD against the data published by New York Metropolitan Transportation Council (NYMTC). Appendix A1 demonstrates a high degree of consistency between the two data sources across travel mode and sector of entry.

We further specify 16 trip segments based on four mutually exclusive population groups, two trip purposes, and two time periods. We define population groups using a hierarchical classification scheme to ensure mutual exclusivity across comparison groups. Individuals aged 65 and older are first identified as seniors, followed by individuals enrolled in school who are classified as students. The remaining population is subsequently divided into low-income and not–low-income groups according to the Federal Poverty Guidelines[2].

[2] https://aspe.hhs.gov/topics/poverty-economic-mobility/poverty-guidelines

Trip purposes are divided into commute and non-commute trips, where commute trips are identified as home-to-work and work-to-home itineraries, and all other trips are categorized as non-commute. Following the CBDTP pricing periods, we distinguish between peak and overnight trips, defining peak hours as 5 a.m.–9 p.m., and treating the remaining hours as overnight. Table 1 lists the number of trips and mode share by segment. Trips made by students to school locations are classified as commute trips. Since seniors may hold part-time jobs, they are also assigned commute trips in Replica's data.

**Table 1**
Number of synthetic trips and driving/transit mode share by segment

| Population group | Trip purpose | Time period | Num. trips (trips/day) | Driving mode share (%) | Transit mode share (%) |
|---|---|---|---|---|---|
| LowIncome | Commute | Peak | 4,492,207 | 33.13% | 28.11% |
| | | Overnight | 427,302 | 30.86% | 30.14% |
| | Non-commute | Peak | 6,146,422 | 40.57% | 8.41% |
| | | Overnight | 1,162,135 | 38.78% | 10.80% |
| NotLowIncome | Commute | Peak | 14,943,566 | 44.25% | 20.42% |
| | | Overnight | 1,264,670 | 41.83% | 22.68% |
| | Non-commute | Peak | 18,199,620 | 50.86% | 4.96% |
| | | Overnight | 3,386,751 | 49.63% | 6.20% |
| Senior | Commute | Peak | 2,457,390 | 48.23% | 14,65% |
| | | Overnight | 313,212 | 46.04% | 14.39% |
| | Non-commute | Peak | 4,037,105 | 54.09% | 2.83% |
| | | Overnight | 827,002 | 52.82% | 3.21% |
| Student | Commute | Peak | 8,937,077 | 18.65% | 5.39% |
| | | Overnight | 239,864 | 40.04% | 17.57% |
| | Non-commute | Peak | 3,411,087 | 48.45% | 4.60% |
| | | Overnight | 619,289 | 47.22% | 5.87% |

*3.1.3 Open map data*

NYMTC defined five sectors of entry for trips entering the CRZ: the 60th Street Sector, Brooklyn Sector, Queens Sector, New Jersey Sector, and Staten Island Sector[3] (NYMTC, 2026). Accordingly, we use Google Maps Directions API to retrieve the shortest route between each county centroid and the CRZ centroid, and based on this determine which entry sector each county's trips pass through to enter the CRZ. If two routes had similar travel times (within 10 minutes), we assign equal weights to their corresponding entry sectors. This allows us to split trips by entry of sectors and compare NYMTC data with Replica data in Appendix A1.

Because Replica's synthetic trip data do not distinguish transit travel time into access, egress, in-vehicle, and wait times, we use OpenTripPlanner (OTP) to obtain this information. OTP is an open-source tool that uses imported Open Street Map (OSM) data for routing on the street network and supports multi-agency public transport routing through imported General Transit Feed Specification (GTFS) data (Young, 2018). We first download OSM data for our study area. Then for the ex- and post-toll periods, we obtain GTFS data for June 2023 and June 2025 from the Mobility Database (Mobility Database, 2026). A matrix of trip origin and destination census block group centroids is generated for retrieval. Transit access time and egress time are calculated using the OSM network distance and a walking speed assumed in OTP. To retrieve transit wait and in-vehicle times, we specify three general departure times in OTP: 9 a.m., 5 p.m., and 10 p.m. For daytime transit travel time ("Peak" period defined in CBDTP), we use the average value of 9 a.m. and 5 p.m. results. For nighttime travel ("Overnight" period defined in CBDTP), we use the 10 p.m. results. Appendix Table A3 presents average transit waiting and in-vehicle times from various trip origin regions to the CRZ.

[3] The Staten Island Sector enters the Manhattan CBD via the Staten Island Ferry service and is considered a separate, clearly defined sector not encompassed in the above other four sectors.

Since OTP-based waiting time is largely determined by one-half the headway between scheduled vehicle arrivals, it may overestimate or underestimated actual waiting time for low-frequency or low-reliability routes. However, we believe this approach remains appropriate for our analysis for two reasons. First, no publicly available dataset provides observed waiting times at a comparable regional scale covering multiple transit agencies across the study area. Collecting observed waiting times from multi-state travelers would be infeasible, and OTP provides a simplified but consistent approach to computing waiting time at this scale. Second, we consider reductions in wait times in the following welfare analysis. Since the half-headway assumption is applied consistently across both scenarios, the impacts of individual travelers checking schedules and service unreliability are captured symmetrically and would not significantly bias the results.

*3.1.4 Vehicle entries, on-demand trips, transit ridership, and toll revenue data*

To capture the changes from 2023 to 2025, we retrieve data from the First Evaluation Report published by the MTA (2026a). Table 2 presents the three key numerical values for post-toll calibration. First, vehicle entries to the CRZ show the baseline (2022-2023 average) and 2025 values for the second quarter, revealing an overall 12.04% reduction in vehicle entries. Second, FHV and taxi trips within the CRZ demonstrate a modest 1.41% increase from 2023 to 2025. Third, transit ridership across all operators serving the CRZ increased by 9.43% from 2023 to 2025, with the MTA accounting for the majority of trips and showing a 9.45% growth in 2025 compared to 2023. Notably, on-demand trips and total transit ridership changed only slightly from 2023 to 2024 but demonstrated substantial shifts from 2023 to 2025. Considering this sudden change in a one-year period, these trends are largely attributed to the implementation of congestion pricing. Calibrating post-toll model ensures that predicted travel behavior accurately reflects the observed modal shifts and traffic reductions following the introduction of congestion pricing.

**Table 2**
Numerical values for calibration retrieved from the MTA first evaluation report

| | 2023 (ex-toll) | 2024 | 2025 (post-toll) | % Change 24 vs. 23 | % Change 25 vs. 23 |
|---|---|---|---|---|---|
| Vehicle entries to CRZ[4] | | | | | |

[4] MTA did not report vehicle entries to CRZ in 2024, the average of 2022 and 2023 is defined as "baseline".

| | | | | | |
|---|---|---|---|---|---|
| April | 644,400 | — | 568,143 | — | -11.83% |
| May | 647,200 | — | 580,226 | — | -10.35% |
| June | 662,400 | — | 570,328 | — | -13.90% |
| Total | 1,954,000 | — | 1,718,697 | — | **-12.04%** |
| FHV & taxi trips in CRZ | | | | | |
| Total | 308,092 | 308,110 | 312,428 | +0.01% | **+1.41%** |
| Transit ridership for all services operating in CRZ | | | | | |
| MTA | 4,671,312 | 4,648,843 | 5,112,676 | -0.48% | +9.45% |
| NJ Transit Corporation | 509,260 | 551,585 | 542,193 | +8.31% | +6.47% |
| NYCDOT | 42,557 | 44,929 | 45,513 | +5.57% | +6.95% |
| NYCEDC | 19,714 | 20,825 | 21,239 | +5.64% | +7.74% |
| Port Authority NY/NJ | 134,793 | 150,931 | 162,694 | +11.97% | +20.70% |
| Roosevelt Island Operating Corporation | 7,084 | 9,354 | 7,936 | +32.05% | +12.03% |
| Total | 5,384,720 | 5,426,467 | 5,892,251 | +0.78% | **+9.43%** |

Note: All values (besides the % change) represent daily average.

According to this report, the toll revenue from January to October is \$573.1 million, and after deducting \$105.3 million in operating expenses, the net revenue is \$467.8 million. To obtain a yearly estimate, we assume that the values in November and December are similar to those in October, resulting in an annual toll revenue of \$704.7 million (\$573.1 + \$65.8×2) and net revenue of \$581.6 million (\$467.8 + \$56.9×2). Moreover, MTA (2025b) reported that 10% of the toll revenue came from trucks, tourist buses, and motorcycles in the first month. Since our models do not consider these vehicle categories, we reduce the revenue estimate by 10% to ensure comparability, resulting in an annual toll revenue of \$634.23 million (\$704.7 × 0.9) and net revenue of \$523.44 million (\$581.6 × 0.9) from resident trips. These values are used to assess whether congestion pricing satisfies Kaldor–Hicks efficiency in Section 4.2. Moreover, we download the number of entries by vehicle class for the second quarter of 2025 from the MTA website (MTA, 2026b). These data are used to validate our post-toll predictions.

### 3.2 Model specification

This section introduces joint mode and destination choice models for both ex- and post-toll periods. Notations used in our models are summarized in Appendix Table A4.

#### *3.2.1 Joint mode and destination model with market-level data*

Given the large trip volume throughout the CSA, we aggregate trips in each segment by origin and destination county and compute the average travel time, average monetary cost, and total number of trips for each mode. The Manhattan county is divided into CRZ and non-CRZ areas to account for the congestion toll. There are two reasons for aggregating the trips to the county level. First, synthetic trips are difficult to validate at the individual level but become more reliable when aggregated into larger spatial units. Second, the county is a suitable geographic unit for congestion pricing studies, as it aligns with common administrative boundaries where policies are implemented and equity concerns are raised. We acknowledge that county-level aggregation does not capture within-county inequities: low-income residents, seniors, and people without good access to transit within the same county may experience the toll burden very differently. This limitation is further discussed in Section 5.1.

We start from the utility function specified in market-level DCMs proposed by Berry et al. (1995). The utility of individual $n \in N$ in market $t \in T$ choosing alternative (or product) $j \in J$ is defined in Eq. (1).

$$U_{njt} = \bar{\delta}_{jt} + \mu_{njt} + \varepsilon_{njt}, \quad \forall n \in N, \forall j \in J, \forall t \in T \tag{1}$$

where $\bar{\delta}_{jt} = x_{jt}\beta - \alpha p_{jt} + \xi_{jt}$ is the systematic utility of alternative $j$ in market $t$; $x_{jt}$ denotes a vector of alternative attributes besides price; $p_{jt}$ denotes the price (monetary cost); $\xi_{jt}$ denotes alternative specific constants; $\beta$ and $\alpha$ are parameters to be estimated. $\mu_{njt}$ denotes the individual-level heterogeneity (e.g., random coefficients interacting with product attributes) utility and $\varepsilon_{njt}$ is a Type I extreme value (Gumbel) distributed random disturbance.

We propose a joint mode and destination model at the market level, in which trips originating from the same county constitute a market indexed by $t$, each mode–destination pair is treated as an alternative indexed by $j$. Moreover, we introduce index $g$ to denote the 16 trip segments defined in Section 3.1.2, with each segment having unique taste parameters. For each market and trip segment pair, we set the individual-level heterogeneity to zero ($\mu_{njt}^{g} = 0, \varepsilon_{njt}^{g} \to \varepsilon_{jt}^{g}$), meaning that trips made by the same population group, with the same purpose, during the same period, and from the same origin, have homogeneous tastes. This assumption simplifies the BLP equation to a standard logit structure that is computationally tractable given a large dataset. After using a compact form to represent terms in $\bar{\delta}_{jt}^{g}$, the utility function can be rewritten as Eq. (2).

$$U_{jt}^{g} = V_{jt}^{g} + \varepsilon_{jt}^{g} = \theta_{g}^{T} X_{jt}^{g} + \varepsilon_{jt}^{g}, \quad \forall j \in J, \forall t \in T, \forall g \in G \tag{2}$$

where $V_{jt}^{g}$ is the systematic utility at the market level; $X_{jt}^{g} = \{x_{jt}^{g}, p_{jt}^{g}, 1\}$ is a vector of alternative attributes, and $\theta_{g} = \{\alpha_{g}, -\beta_{g}, \xi_{g,jt}\}$ is a vector of taste parameters. Market-level models typically include an outside alternative ($j = 0$), representing the option of "not buying anything" (Berry, 1994). We treat traveling to destinations outside the CSA, regardless of travel mode, as the outside alternative and use a market-specific constant to represent its systematic utility ($V_{0t}^{g} = o_{t}^{g}$).

The market share of alternative $j$ in market $t$ and segment $g$ is given by Eq. (3), and its logarithmic ratio relative to the outside alternative is presented in Eq. (4).

$$s_{jt}^{g} = \frac{e^{V_{jt}^{g}}}{\sum_{q \in J} e^{V_{qt}^{g}}}, \quad \forall j \in J, \forall t \in T, \forall g \in G \tag{3}$$

$$\ln\left(s_{jt}^{g}/s_{0t}^{g}\right) = \ln\left(e^{V_{jt}^{g}}/e^{V_{0t}^{g}}\right) = V_{jt}^{g} - o_{t}^{g}, \quad \forall j \in J, j \neq 0, \forall t \in T, \forall g \in G \tag{4}$$

where $s_{jt}^{g}$ denotes the market share (choice probability) of alternative $j$ in market $t$ and segment $j$. The logarithm form of the ratio between the market share of alternative $j$ and that of the outside alternative is called the inverse market share of $j$ (Berry, 1994).

Our study area covers 31 spatial units (29 counties plus the CRZ and upper Manhattan). Six travel modes are considered: driving, public transit, fhv (including taxi and for-hire vehicles), biking, walking, and carpool (trips made by multiple passengers in a single auto vehicle). Accordingly, each trip segment has 31 markets and $31 \times 6 = 186$ alternatives. Since each alternative is a combination of travel mode and destination, its systematic utility $V_{jt}^{g}$, where $j = (m, d)$, can be decomposed by mode $m$ and destination $d$. We include alternative attributes such as travel time, monetary cost, mode-specific constants, county-specific constants, and region-related interaction terms. The systematic utilities of the six modes for any $t \in T, d \in D, g \in G$ are specified in Eqs. (5) – (10).

$$\begin{aligned} V_{driving,d,t}^{g} = {} & \left(\theta_{g,autoTT} + \theta_{g,autoTT}^{NYC} IsNYC_t\right) TT_{driving,d,t}^{g} \\ & + \left(\theta_{g,cost} + \theta_{g,cost}^{NYC} IsNYC_t\right) CO_{driving,d,t}^{g} + \theta_{g,asc}^{driving} + \theta_{g,asc}^{d} \end{aligned} \tag{5}$$

$$\begin{aligned} V_{transit,d,t}^{g} = {} & \left(\theta_{g,AT} + \theta_{g,AT}^{NYC} IsNYC_t\right) AT_{transit,d,t}^{g} + \left(\theta_{g,ET} + \theta_{g,ET}^{NYC} IsNYC_t\right) ET_{transit,d,t}^{g} \\ & + \left(\theta_{g,WT} + \theta_{g,WT}^{NYC} IsNYC_t\right) WT_{transit,d,t}^{g} \\ & + \left(\theta_{g,IVT} + \theta_{g,IVT}^{NYC} IsNYC_t\right) IVT_{transit,d,t}^{g} + \theta_{g,trans} Trans_{transit,d,t}^{g} \\ & + \left(\theta_{g,cost} + \theta_{g,cost}^{NYC} IsNYC_t\right) CO_{transit,d,t}^{g} + \theta_{g,asc}^{transit} + \theta_{g,asc}^{d} \end{aligned} \tag{6}$$

$$\begin{aligned} V_{fhv,d,t}^{g} = {} & \left(\theta_{g,autoTT} + \theta_{g,autoTT}^{NYC} IsNYC_t\right) TT_{fhv,d,t}^{g} + \left(\theta_{g,cost} + \theta_{g,cost}^{NYC} IsNYC_t\right) CO_{fhv,d,t}^{g} \\ & + \theta_{g,asc}^{fhv} + \theta_{g,asc}^{d} \end{aligned} \tag{7}$$

$$V_{biking,d,t}^{g} = \left(\theta_{g,nonautoTT} + \theta_{g,nonautoTT}^{NYC} IsNYC_t\right) TT_{biking,d,t}^{g} + \theta_{g,asc}^{biking} + \theta_{g,asc}^{d} \tag{8}$$

$$V_{walking,d,t}^{g} = \left(\theta_{g,nonautoTT} + \theta_{g,nonautoTT}^{NYC} IsNYC_t\right) TT_{walking,d,t}^{g} + \theta_{g,asc}^{walking} + \theta_{g,asc}^{d} \tag{9}$$

$$V_{carpool,d,t}^{g} = \left(\theta_{g,autoTT} + \theta_{g,autoTT}^{NYC} IsNYC_t\right) TT_{carpool,t}^{g} + \theta_{g,asc}^{d} \tag{10}$$

where $TT_{m,d,t}^{g}$ and $CO_{m,d,t}^{g}$ are the average travel time (in minutes) and monetary cost (in dollars); $AT_{transit,d,t}^{g}$, $ET_{transit,d,t}^{g}$, $WT_{transit,d,t}^{g}$, $IVT_{transit,d,t}^{g}$, $Trans_{transit,d,t}^{g}$ are the access time, egress time, wait time, in-vehicle time, and number of transfers for taking public transit. $IsNYC_t$ is a binary variable that equals 1 if market $t$ is located within NYC and 0 otherwise. This specification follows a common approach in the literature, as many studies treat NYC as a distinct region due to its high transit dependency and different sensitivity patterns to travel time and cost compared to other U.S. regions (Hwang et al., 2015; Salon, 2009).We use destination-specific constants rather than attributes, such as facility proximity or employment density, since these attributes are not directly affected by congestion pricing or its compensatory transit strategies. All terms associated with $\theta$ represent taste parameters to be estimated, including sensitivities to travel time and cost as well as mode- and destination-specific constants.

### *3.2.2 Pre-implementation model estimation*

In the joint mode and destination choice model, the random disturbances are not assumed to be i.i.d.; rather, they exhibit correlation both across modes to the same destination and across destinations when choosing the same mode. Previous studies typically address one of these correlation structures by estimating nested logit (NL) models, capturing dependence either across destinations with upper-branch destination choice and lower-branch mode nests, or the reverse (Bills et al., 2022; Newman & Bernardin, 2010). However, this is at the risk of misunderstanding substitution and complementarity patterns (Huo et al., 2024).

Fosgerau et al. (2024) proposed inverse product differentiation logit (IPDL) to address the limitation of hierarchical structures and provide faster estimation. IPDL allows alternatives to be nested across multiple hierarchical dimensions $h \in H$, with each alternative belonging to exactly one nest within each dimension. Huo et al., (2024) proved that IPDL is a general form of multinomial logit (MNL) and nested logit (NL). MNL is obtained when there is no hierarchical structure ($H = 0$). NL is obtained when there is only one hierarchical structure ($H = 1$). Fig. 2 illustrates a simplified case showing the difference between the NL and the IPDL in mode choice prediction. In the NL model, a reduction in driving within the CRZ results in a general increase for all modes in the non-CRZ area, since the two branches are independent. By contrast, the IPDL model captures cross-dimensional substitution, where a reduction in driving within the CRZ may also lead to a decline in driving outside the CRZ, as driving overall becomes less attractive. This enables us to capture the broader spillover effects of congestion tolls.

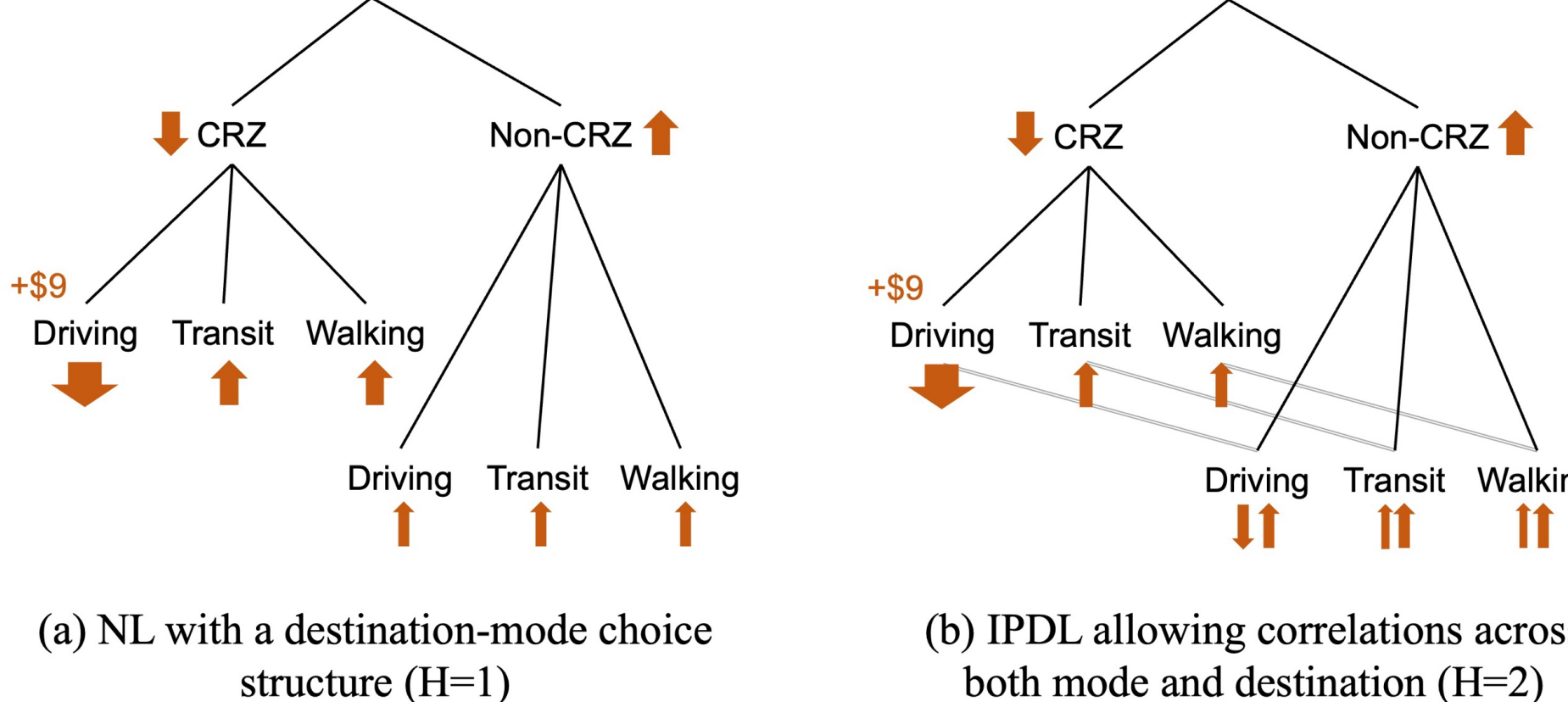

(a) NL with a destination-mode choice structure (H=1)

(b) IPDL allowing correlations across both mode and destination (H=2)

**Fig. 2.** A comparison between NL and IPDL.

We estimate a separate model for each trip segment, and remove the index $g$ in this section for simplicity. For each segment $g$, IPDL directly specifies the market-level systematic utility as Eqs. (11) – (12).

$$\bar{\delta}_{jt} = V_{jt} = \ln G_j(s_t; \varphi) + c_t, \forall j \in J, \forall t \in T \tag{11}$$

$$\ln G_j(s_t; \varphi) = \left(1 - \sum_{h=1}^{H} \rho_h\right) \ln(s_{jt}) + \sum_{h=1}^{H} \rho_h \ln\left(\sum_{q \in J_h} s_{qt}\right), \forall j \in J, \forall t \in T \tag{12}$$

where $G_j(s_t; \varphi)$ is the invertible function of market share $s_t$, $\varphi$ is a vector of parameters, $c_t$ is a constant for market $t$, and $J_h$ is a set of alternatives grouped by dimension $h$. The higher value of $\rho_h$ implies that the error terms of alternatives ($\varepsilon_{jt}$) in the group classified in dimension $h$ have stronger correlations. To this end, correlation among multiple dimensions is captured by $\varphi = \{\rho_1, \rho_2, \dots, \rho_H\}$. Since the systematic utility of the outside alternative is set as a market-specific constant ($V_{0t} = o_t$), we have $V_{0t} = \ln G_j(s_t; \varphi) + c_t = \ln(s_{0t}) + c_t = o_t \rightarrow c_t = o_t - \ln(s_{0t})$. Linking this to Eqs. (11) – (12), we obtain Eq. (13) that relates the inverse market share to alternative attributes and nesting variables. Details for the derivation of Eq. (13) can be found in Appendix A4.

$$\ln\left(\frac{s_{jt}}{s_{0t}}\right) = \theta^T X_{jt} + \sum_{h=1}^{H} \rho_h \ln\left(\frac{s_{jt}}{\sum_{q\in J_h} s_{qt}}\right) - o_t, \qquad \forall j \in J, j \neq 0, \forall t \in T \tag{13}$$

where $\ln\left(\frac{s_{jt}}{\sum_{q\in J_h} s_{qt}}\right)$ serves as a nesting variable associated with hierarchical dimension $h$, with parameter $\rho_h$ to be estimated. Fosgerau et al. (2024) demonstrated that estimating IPDL reduces to a linear regression, where the dependent variable is the log-ratio of market shares $\ln(s_{jt}/s_{0t})$, and the independent variables are the observed alternative attributes $X_{jt}$ along with the nesting variables $\ln\left(\frac{s_{jt}}{\sum_{q\in J_h} s_{qt}}\right)$ for each grouping dimension. The market-specific constant $o_t$ is absorbed by market fixed effects in estimation. This makes IPDL a more computationally efficient and scalable approach for estimating demand in settings with large choice sets. Since Eq. (13) holds for all alternatives and markets, the total number of regression observations is $|J| \times |T|$.

IPDL can be estimated using the two-stage least squares (2SLS) approach to handle endogeneity bias in market-level models (Angrist & Krueger, 2001). Following Krueger et al. (2023)'s and Ren et al. (2025)'s work, we treat travel cost as an endogenous variable. We introduce BLP-style instruments: monetary cost of other alternatives in the same origin, destination, and mode groups are averaged separately, resulting in creating three instrumental variables (IVs) per model. For each trip segment $g \in G$, we run instrumental regression to estimate unbiased taste parameters $\theta_g$, serving as the foundation for welfare analysis.

*3.2.3 Post-implementation parameter calibration*

Since limited data is available for the post-toll period, it is not feasible to estimate a complete choice model with confidence intervals. Instead, we introduce toll-related parameters to capture the general effects of congestion pricing on travelers' preferences. Additionally, we incorporate several changes in alternative attributes. First, congestion tolls are applied to auto trips entering the CRZ. Second, according to empirically observed data from Cook et al. (2025), there is a 15% increase in average driving speeds within the CRZ. To reflect this improvement, we split each county-level OD path (the line linking county centroids) using the geographic boundary of the CRZ, calculate the proportion of the path located within the zone, and apply a corresponding 15% reduction in auto travel time for that segment. Third, we account for changes in transit fare and service performance between 2023 and 2025, which are captured through OTP, with selected results presented in Appendix Table A3. Fourth, we consider fuel price changes over the study period. According to Federal Reserve Economic Data (FRED)[5], the average fuel price in the New York-Newark-Jersey City metropolitan area decreased from \$3.47 per gallon in Q2 2023 to \$2.99 per gallon in Q2 2025, a decline of 13.83%. We apply this same reduction to both driving and carpool costs. Together, these adjustments result in the utility function for the post-toll period as shown in Eqs. (14) – (16).

$$V_{jt}^{post,g} = \theta_{g,j}^{post^T} X_{jt}^{post,g}, \qquad \forall j \in J, \forall t \in T, \forall g \in G \tag{14}$$

[5] https://fred.stlouisfed.org/series/APUS12A74714

$$X_{jt}^{post,g} = X_{jt}^{ex,g} + \Delta^{toll} X_{ij}^{g} + \Delta^{other} X_{ij}^{g}, \qquad \forall j \in J, \forall t \in T,, \forall g \in G \tag{15}$$

$$\begin{aligned} \theta_{g,j}^{post} &= \theta_{g}^{ex} + \Delta^{toll}\theta_{g,j} \\ &= \theta_{g}^{ex} + \theta_{asc-toll}^{CRZ} + \sum_{m\prime \in M_{toll}} \theta_{asc-toll}^{m\prime} \cdot 1[m = m'] \cdot 1[d = CRZ], \\ &\forall j = (m, d), \forall g \in G \end{aligned} \tag{16}$$

where $X_{jt}^{ex,g}$ and $\theta_{g}^{ex}$ denote the alternative attributes and taste parameters in the ex-toll model; $\Delta^{toll} X_{ij}^{g}$ represents changes in alternative attributes directly resulting from the congestion toll, specifically the toll charges applied to auto trips and the associated travel time reductions due to improved driving speeds; and $\Delta^{other} X_{ij}^{g}$ captures other attribute changes unrelated to the toll, including updates to transit fare and service performance and fuel price reductions. The post-toll parameter $\theta_{g,j}^{post}$ is indexed by alternative $j = (m, d)$, as the effects of congestion pricing on traveler preferences vary across travel modes and destinations. $\theta_{asc-toll}^{CRZ}$ reflects the general effect on the attractiveness of CRZ as a destination. $\theta_{asc-toll}^{m\prime}$ reflects the effects on mode attractiveness for trips entering the CRZ. $M_{toll} = \{driving, transit, fhv, carpool\}$ is a set of modes potentially impacted by the program.

Calibrating the post-toll model presents an underdetermined system: we seek to adjust five constants ($\theta_{asc-toll}^{driving}$, $\theta_{asc-toll}^{transit}$, $\theta_{asc-toll}^{fhv}$, $\theta_{asc-toll}^{carpool}$, and $\theta_{asc-toll}^{CRZ}$) to match three observed percentage changes in vehicle entries, on-demand trips, and transit ridership mentioned in Section 3.1.4. To obtain a unique and reasonable solution, we employ the Moore-Penrose pseudoinverse approach (Moore, 1920), which identifies the minimum-norm solution—the smallest adjustment to the taste parameters that satisfies the calibration targets. This approach embodies the principle of parsimony: among all feasible calibrations, we select the one requiring the least departure from the original model specification.

We implement the calibration by first linearizing the relationship between the coefficient adjustments and the outcome metrics via finite difference approximation of the Jacobian matrix, which captures the sensitivity of each percentage change to each toll-related parameter. The linearized system is then solved using the pseudoinverse to obtain the minimum-norm solution. To account for nonlinearities in the underlying discrete choice model, we iterate this procedure—updating the Jacobian at each step—until the residual between predicted and target outcomes falls below a convergence tolerance of $10^{-8}$.

### 3.3 Metrics for welfare analysis

Based on the taste parameters from the choice models, we calculate several metrics for welfare analysis, including value of time, consumer surplus, and compensating variation.

Value of time (VOT) measures traveler's trade-off between time savings and monetary costs, reflecting their willingness to pay for reduced travel time (Small, 2012). Following existing studies, we compute VOTs as the marginal rate of substitution between travel time and travel cost. For instance, the value of auto travel time for trip segment $g \in G$ is defined in Eq. (17).

$$VOT_{auto_tt}^{g} = \frac{\theta_{g,autoTT}}{\theta_{g,cost}}, \qquad \forall g \in G \tag{17}$$

where $G$ is a set containing 16 trip segments; $\theta_{g,autoTT}$ is the parameter of auto travel time for segment $g$; $\theta_{g,cost}$ is the cost parameter for segment $g$. To quantify the precision of VOT estimates, we report standard errors computed using the normal approximation approach (Armstrong et al., 2001), as shown in Eq. (18).

$$S.E.(VOT) = \frac{1}{|\theta_{cost}|}\sqrt{\sigma_{time}^{2} + \frac{\theta_{time}}{\theta_{cost}}\left(\frac{\theta_{time}}{\theta_{cost}}.\sigma_{cost}^{2} - 2\rho\sigma_{time}\sigma_{cost}\right)} \tag{18}$$

where $\theta_{time}$ and $\theta_{cost}$ are corresponding time and cost parameters from the same segment, $\sigma_{time}$ and $\sigma_{cost}$ are their standard errors, and $\rho$ is their correlation. Because the variance–covariance matrix is not considered in our model, we set $\rho = 0$, which yields conservative standard errors.

Consumer surplus (CS) is an economic concept that quantifies consumer welfare using the difference between the highest price a consumer is willing to pay for a good or service and the actual price they pay (Small & Rosen, 1981). Consistent with other choice models, CS in the IPDL framework can be computed as the logsum of alternative utilities (Fosgerau et al., 2024), as shown in Eq. (19).

$$CS_{t}^{g} = \ln\left(\sum_{j\in J} H_{j}^{g}\left(e^{V_{jt}^{g}}\right)\right) + C^{g} = c_{t}^{g} + C^{g} = -\ln\left(s_{0t}^{g}\right) + C^{g}, \forall t \in T, g \in G \tag{19}$$

where $H_{j}^{g}\left(e^{V_{jt}^{g}}\right) = {G^{g}}_{j}^{-1}\left(e^{V_{jt}^{g}}\right)$ denotes the utility function adjusted by considering alternative correlations, $c_{t}^{g}$ is the constant for market $t$ of segment $g$, and $C^{g}$ is an unknown constant.

CSs from different model specifications cannot be directly compared due to the unknown constant. However, changes in CSs can be converted to monetary units, and thus comparable units using compensating variation (CV), which is interpreted as the dollar amount an individual would have to be compensated to be as well off as before a policy change (Freeman et al., 2014). In our study, the total CV brought by changes from 2023 to 2025 is defined in Eq. (20).

$$CV_{t}^{g} = -\frac{1}{\theta_{g,cost}}\left(CS_{t}^{post,g} - CS_{t}^{ex,g}\right)d_{t}^{g}, \qquad \forall t \in T, g \in G \tag{20}$$

where $\theta_{g,cost}$ is the cost parameter for segment $g$, serving as a proxy for the marginal utility of income. $CS_{t}^{ex,g}$ and $CS_{t}^{post,g}$ denote consumer surplus in 2023 and 2025, while $d_{t}^{g}$ denotes the total trip demand made by segment $g$ in market $t$.

Since the total CV includes toll-related and non-toll-related impacts, we need to isolate the welfare effects attributable to congestion pricing. We first express the systematic utility in 2025 as a function of the 2023 baseline and the changes that occurred, as shown in Eq. (21).

$$V_{jt}^{post,g} = \left(\theta_g^{ex} + \Delta^{toll}\theta_{g,j}\right)^T \left(X_{jt}^{ex,g} + \Delta^{toll}X_{ij}^g + \Delta^{other}X_{ij}^g\right) \\ = V_{jt}^{ex,g} + A + B + C + D + E, \qquad \forall j \in J, t \in T, g \in G \tag{21}$$

where all the elements are defined in Eqs. (14) – (16). We defined five utility components as: $A = \theta_g^{ex}\Delta^{toll}X_{ij}^g$, $B = \theta_g^{ex}\Delta^{other}X_{ij}^g$, $C = \Delta^{toll}\theta_{g,j}X_{jt}^{ex,g}$, $D = \Delta^{toll}\theta_{g,j}\Delta^{toll}X_{ij}^g$, and $E = \Delta^{toll}\theta_{g,j}\Delta^{other}X_{ij}^g$. Table 3 summarizes these components. We attribute a component to the congestion toll if it involves either the toll-induced preference shift ($\Delta^{toll}\theta_{g,j}$) or the toll-related attribute change $\Delta^{toll}X_{ij}^g$). Under this rule, only Component B is classified as non-toll.

**Table 3**
Decomposed utility components and their interpretation

| | Expression | Interpretation | Attribution |
|---|---|---|---|
| A | $\theta_g^{ex}\Delta^{toll}X_{ij}^g$ | Direct toll attribute effect evaluated at 2023 preferences | Toll |
| B | $\theta_g^{ex}\Delta^{other}X_{ij}^g$ | Non-toll attribute changes evaluated at 2023 preferences | Non-toll |
| C | $\Delta^{toll}\theta_{g,j}X_{jt}^{ex,g}$ | Interaction: toll-induced preference shift × baseline attributes | Toll |
| D | $\Delta^{toll}\theta_{g,j}\Delta^{toll}X_{ij}^g$ | Interaction: toll-induced preference shift × toll attribute changes | Toll |
| E | $\Delta^{toll}\theta_{g,j}\Delta^{other}X_{ij}^g$ | Interaction: toll-induced preference shift × non-toll attribute changes | Toll |

Correctly measuring the welfare impact of congestion pricing requires isolating the contribution of Component B (non-toll attribute changes) from the toll-related components. A single counterfactual comparison cannot achieve this because removing the toll-related attribute changes ($\Delta^{toll}X_{ij}^g$) alone, removing the preference shift ($\Delta^{toll}\theta_{g,j}$) alone, or both simultaneously each yields a different result, and none of them separates Component B from the toll-related effects. To address this issue, we employ the Shapley value decomposition (Shapley, 1953), which is adopted in economic analysis for allocating total welfare changes among contributing factors when interactions between factors create path-dependence (Shorrocks, 2013). For any $t \in T$ and $g \in G$, the Shapley value decomposed CV for component $k$ is defined in Eqs. (22) – (23).

$$\omega_k = \sum_{S \subseteq N/\{k\}} \frac{|S|!\,(|K| - |S| - 1)!}{|K|!} \times [cv(S \cup \{k\}) - cv(S)], \quad \forall k \in K \tag{22}$$

$$cv(S) = -\frac{1}{\theta_{cost}} \left[ \ln \sum_{j \in J} \exp\left(V_j^{ex} + \sum_{k \in S} V_j^k\right) - \ln \sum_{j \in J} \exp\left(V_j^{ex}\right) \right] d, \tag{23}$$

where $K$ is a set of all five components, $S$ is a set of components of interest, $V_j^k$ denotes the change in alternative $j$'s utility brought by component $k$, and $cv(S)$ represents the compensating variation (in monetary terms) when only components in set $S$ are "switched on." Based on this formulation, the CV attributable to congestion pricing for market $t \in T$ and segment $g \in G$ is defined in Eq. (24).

$$CV_t^{toll,g} = \omega_{A,t}^g + \omega_{C,t}^g + \omega_{D,t}^g + \omega_{E,t}^g, \qquad \forall t \in T, g \in G \tag{24}$$

### 3.4 Compensatory transit strategies and experimental design

The allocation of congestion toll revenues toward transit improvements is critical to the long-term viability and public acceptance of NYC's congestion pricing program. Because our modeling framework captures travel behavior through mode and destination choice, it provides a natural basis for evaluating how compensatory transit strategies can offset potential welfare losses induced by tolling. Since post-implementation data are limited, we rely on our joint mode–destination choice framework to provide a principled and internally consistent basis for counterfactual welfare analysis.

We evaluate two compensatory transit strategies: reducing wait time and providing fare discounts. In practice, transit improvements can be achieved through multiple operational strategies, including increasing service frequency, improving schedule reliability, and optimizing transfer coordination between routes (Osuna & Newell, 1972). Each of these interventions ultimately reduces the effective wait time and overall travel disutility experienced by passengers. We therefore treat wait time reduction as a composite proxy metric that captures the effect of a range of service improvements, rather than prescribing a single operational lever whose cost is difficult to measure and falls outside the scope of this research. Transit fare discount, on the other hand, directly alleviates the monetary burden on travelers, which is adopted as a short-term strategy to enhance affordability and promote public transit usage, especially when pricing policies increase out-of-pocket travel costs (Paulley et al., 2006). We operationalize these strategies by applying reductions in average waiting times and population-specific fare discounts to corresponding transit trips.

These compensatory strategies are evaluated under two distinct objectives: achieving Kaldor–Hicks efficiency (Kaldor, 1939) and ensuring Pareto improvement (Varian, 1992). The former emphasizes whether the aggregate gains to the winners outweigh the aggregate losses to the losers, while the latter requires that no identifiable group is made worse off. Under the Kaldor–Hicks efficiency scenario, we evaluate the two strategies independently and calculate the amounts of wait time reduction or fare discount required to offset the aggregate welfare loss, as shown in Eqs. (25) – (26).

$$WTC^{KH} = f_{wt}\left(\sum_{g \in G} \sum_{t \in T} CV_t^{toll,g}\right) \tag{25}$$

$$FDC^{KH} = f_{cost}\left(\sum_{t \in T_c} CV_t\right) \tag{26}$$

where $\sum_{g \in G} \sum_{t \in T} CV_t^{toll,g}$ represents the total CV related to congestion pricing. $f_{wt}(\cdot)$ and $f_{cost}(\cdot)$ are functions that take CV as inputs and return the corresponding amounts of wait time and fare discount compensation needed to offset those changes. Since CS changes nonlinearly with respect to alternative attributes, these functions do not have closed-form solutions. Instead, we use a numerical root-finding approach to identify the values of $WTC^{KH}$ or $FDC^{KH}$ that reproduce the CV induced by the congestion toll. This is implemented as an optimization problem where $WTC^{KH}$ or $FDC^{KH}$ is defined as the decision variable, and the objective is to minimize the squared difference between the given and reproduced CVs, solved using the Sequential Least Squares Programming (SLSQP) algorithm in SciPy.

Under the Pareto improvement scenario, a single uniform compensation measure is generally insufficient, as traveler groups differ in their values of time, trip patterns, and sensitivity to monetary costs. Ideally, compensation would be tailored at the individual trip level to ensure that no single trip is made worse off. However, such granularity is infeasible from a policy implementation standpoint: transit agencies cannot design distinct fare discounts for each individual trip. We therefore define compensation at the population group level, where each group $i \in I = \{NotLowIncome, LowIncome, Senior, Student\}$ receives a uniform fare discount. This group-level design strikes a pragmatic balance between welfare equity and operational feasibility. Specifically, we adopt a combined strategy in which welfare losses are first partially offset through incremental reductions in wait time, applied at 5% intervals. For each level of wait time reduction, we then calculate the remaining uncompensated welfare, which is subsequently addressed through population-specific fare discounts. This sequential approach reflects a realistic policy design in which system-wide service improvements are complemented by targeted fare subsidies. We report the percentage decreases in wait time (%) and the corresponding per-trip subsidies (in dollars) needed to achieve full welfare compensation. Given a 5% reduction in wait time, the fare discount for population $i$ to ensure Pareto improvement ($FDC^{Pareto,i}$), is defined in Eqs. (27) – (29).

$$CV_t^{remain,g} = \min(CV_t^{toll,g} + CV_t^{5\%,g}, 0), \qquad \forall t \in T, \forall g \in G \tag{27}$$

$$FDC_t^{Pareto,i} = f_{cost}\left(\sum_g CV_{t \in G_i}^{remain,g}\right), \qquad \forall t \in T, \forall i \in I \tag{28}$$

$$FDC^{Pareto,i} = \max_{t \in T} FDC_t^{Pareto}, \qquad \forall i \in I \tag{29}$$

where $CV_t^{5\%,g}$ represents the CV brought by 5% reduction in wait time, $CV_t^{remain}$ represents the remaining welfare loss (capped at zero to retain only uncompensated losses) after the wait time improvement has been applied. The set $G_i$ contains all trip segments belonging to population group $i$. Eq. (28) ensures that the fare discount for each population group in each market is sufficient to compensate the group's total remaining CV across trip purposes and time periods. Eq. (29) ensures that the fare discount for each population group covers the largest welfare loss among all origin counties, thereby achieving group-level Pareto improvement. We note that the transit improvements and fare discounts themselves may induce further changes in mode and destination choices; accordingly, we use our model to predict transit ridership under each compensation scenario, so that the endogenous mode and destination shifts induced by the compensation policy can be explicitly measured.

The proposed approach provides a reasonable first-order approximation of compensation requirements when complete post-implementation data on transit improvements are not yet available. As empirical data from the congestion pricing program accumulate over time, these model-based estimates can be validated and refined, but they offer actionable policy guidance in the interim. In addition, the actual cost of achieving specific wait time reductions depends on agency-specific factors (fleet availability, labor costs, infrastructure constraints) that vary across operators. Nevertheless, our analysis quantifies the accessibility-based welfare compensation achieved at each level of wait time reduction, providing transit agencies with a roadmap for investment prioritization.

# 4. Results

This section presents the results of choice models, welfare analysis, and transit policy evaluations. The experiments were conducted on a local machine equipped with an Intel Core i7-10875H CPU and 32GB of RAM. The AER package in R was used for MNL, NL, and IPDL estimations, while the remaining analyses were performed in Python.

## 4.1 Estimated choice models

We first validate the choice of our model specification over simpler model structures. Appendix Table A5 compares the MNL, NL, and IPDL specifications estimated on total trips without segmentation. The IPDL achieves a substantially lower NMAE (10.08%) compared to the NL (32.67%) and MNL (73.77%), confirming the importance of capturing substitution patterns across both mode and destination dimensions.

Appendix Table A6 further examines the sensitivity of the IPDL to its key components: fixed effects for outside alternative utility, NYC-specific interaction terms, and instrumental variables. Removing market-level fixed effects or interaction terms increases the NMAE to 16.64% and 18.62%, respectively, while the version without instrumental variables shows only a marginal improvement in NMAE (9.47% vs. 10.08%), indicating small prediction loss from the endogeneity correction. Based on these results, we adopt the full IPDL specification with instrumental variables, market fixed effects, and interaction terms for all subsequent analyses.

### *4.1.1 Basic statistics*

Table 4 summarizes the IPDL estimates of ex-toll models for four select trip segments. The reported values include mean estimates, standard errors, significance levels, and meta information. For brevity, only the destination-specific constant for the CRZ is reported. Model parameters for other trip segments are provided in Appendix Tables A8–A11. In general, the normalized mean absolute error (NMAE) remains below 5% for most segment models, indicating strong predictive accuracy. Most parameters are significant, and the estimates align with existing studies on mode and destination choice (He et al., 2021).

Transit wait time penalties are significant for non-low-income commuters but become insignificant for low-income commuters, seniors, and students, suggesting that non-commute trips made by these populations may exhibit greater tolerance toward delayed schedules. In contrast, transfer penalties show the opposite pattern: they are not significant for non-low-income commuters but become significant for low-income commuters, seniors, and students.

The significant nesting parameters confirm the appropriateness of the IPDL framework: travelers exhibit substitution patterns both across modes serving the same destination and across destinations accessible by the same mode.

Mode- and destination-specific constants provide insight into baseline preferences. Driving constants are positive and significant across segments, underscoring the relative attractiveness of private auto use, while biking and walking constants are strongly negative, reflecting limited substitution toward active modes. The transit constant is positive and significant for commuters but becomes negative or insignificant for seniors and students, suggesting that transit is a more attractive option for commute trips. The CRZ-specific constants are not statistically significant for most segments, indicating no strong baseline preference or aversion toward traveling to this area after controlling for travel time and cost.

NYC-specific interaction terms highlight spatial heterogeneity in preferences. For example, auto travel time and transit in-vehicle time in New York City are more negatively perceived compared to other regions, suggesting that travelers in NYC often operate under more rigid time schedules. Conversely, cost interactions for trips destined to NYC show positive values across all groups, potentially indicating a higher baseline willingness to pay among travelers in NYC.

**Table 4**
Ex-toll model results for selected trip segments (each entry represents the average value, and the number in the parenthesis is the standard error).

| | **NotLowIncome, Commute, Peak** | **LowIncome, Commute, Peak** | **Senior, Non-commute, Peak** | **Student, Non-commute, Peak** |
|---|---|---|---|---|
| **Travel time and cost** | | | | |
| Auto travel time ($\theta_{autoTT}$) | -0.106*** (0.007) | -0.098*** (0.008) | -0.036*** (0.007) | -0.038*** (0.008) |
| Transit access time ($\theta_{AT}$) | -0.157*** (0.019) | -0.202*** (0.016) | -0.138** (0.046) | -0.074*** (0.019) |
| Transit egress time ($\theta_{ET}$) | -0.179*** (0.014) | -0.163*** (0.014) | -0.142* (0.060) | -0.079*** (0.020) |
| Transit wait time ($\theta_{WT}$) | -1.423* (0.725) | -1.659 (3.784) | -0.588 (1.888) | -0.575 (0.753) |
| Transit in-vehicle time ($\theta_{IVT}$) | -0.083*** (0.009) | -0.033*** (0.009) | 0.026 (0.078) | -0.017 (0.065) |
| Number of transfers ($\theta_{trans}$) | -0.040 (0.054) | -0.166** (0.055) | -0.637*** (0.117) | -0.555*** (0.117) |
| Non-vehicle travel time ($\theta_{nonautoTT}$) | -0.062*** (0.007) | -0.039*** (0.006) | 0.023 (0.023) | 0.039 (0.023) |
| Trip cost ($\theta_{cost}$) | -0.277*** (0.024) | -0.460*** (0.079) | -0.443*** (0.033) | -0.470*** (0.040) |
| **NYC-specific interaction terms** | | | | |
| Auto travel time ($\theta_{autoTT}^{NYC}$) | -0.019*** (0.002) | -0.022*** (0.002) | -0.026*** (0.002) | -0.033*** (0.002) |
| Transit access time ($\theta_{AT}^{NYC}$) | -0.075*** (0.010) | -0.013*** (0.001) | -0.009*** (0.002) | -0.020*** (0.004) |
| Transit egress time ($\theta_{ET}^{NYC}$) | 0.003 (0.011) | 0.012 (0.011) | -0.011 (0.052) | -0.027 (0.032) |
| Transit wait time ($\theta_{WT}^{NYC}$) | -0.177 (0.181) | 0.210 (0.181) | -0.473 (0.808) | -0.695 (0.654) |
| Transit in-vehicle time ($\theta_{IVT}^{NYC}$) | -0.025*** (0.004) | -0.015*** (0.004) | -0.026 (0.017) | -0.024 (0.014) |
| Non-vehicle travel time ($\theta_{nonautoTT}^{NYC}$) | -0.008*** (0.000) | -0.005*** (0.000) | -0.053*** (0.003) | -0.050*** (0.003) |
| Trip cost ($\theta_{cost}^{NYC}$) | 0.087*** (0.006) | 0.073*** (0.010) | 0.106*** (0.007) | 0.113*** (0.008) |
| **Mode and destination constant** | | | | |
| Driving constant ($\theta_{asc}^{driving}$) | 0.339*** (0.011) | 0.197*** (0.011) | 0.376*** (0.011) | 0.365*** (0.011) |
| Transit constant | 1.040*** | 1.017*** | -0.205* | 0.095 |
| ($\theta_{asc}^{transit}$) | (0.048) | (0.050) | (0.102) | (0.093) |
| FHV constant | -1.183*** | -1.562*** | -0.728*** | -0.875*** |
| ($\theta_{asc}^{fhv}$) | (0.070) | (0.087) | (0.043) | (0.051) |
| Biking constant | -2.004*** | -1.853*** | -1.759*** | -1.625*** |
| ($\theta_{asc}^{biking}$) | (0.066) | (0.061) | (0.057) | (0.052) |
| Walking constant | -0.606*** | -0.440*** | -0.423*** | -0.232*** |
| ($\theta_{asc}^{walking}$) | (0.029) | (0.027) | (0.027) | (0.024) |
| CRZ-specific constant | 0.152 | 0.066 | -0.058 | 0.071 |
| ($\theta_{asc}^{CRZ}$) | (0.129) | (0.137) | (0.111) | (0.154) |
| **Nesting parameter** | | | | |
| $\ln\left(\frac{s_{jt}}{\sum_{q\in J_{mode}} s_{qt}}\right)$ | 0.292***<br>(0.002) | 0.299***<br>(0.002) | 0.324***<br>(0.002) | 0.327***<br>(0.002) |
| $\ln\left(\frac{s_{jt}}{\sum_{q\in J_{destination}} s_{qt}}\right)$ | 0.161***<br>(0.003) | 0.167***<br>(0.003) | 0.183***<br>(0.003) | 0.175***<br>(0.003) |
| **Meta information** | | | | |
| Instrumental variables | Yes | Yes | Yes | Yes |
| Market fixed effects | Yes | Yes | Yes | Yes |
| # Observations | 3,291 | 3,042 | 2,943 | 2,838 |
| Total trips per day | 14,943,566 | 4,492,207 | 4,037,105 | 3,411,087 |
| MAE (num. trips) | 191.07 | 66.60 | 29.54 | 46.49 |
| NMAE (%) | 4.21% | 4.51% | 2.15% | 3.87% |

Note: ***p-value<0.001, **p-value<0.01, *p-value<0.05. MAE denotes the mean absolute error of alternative-level trip volume predictions, and NMAE denotes the normalized MAE, obtained by scaling MAE by the mean trip volume.

### *4.1.2 Model prediction and validation*

Fig. 3 presents predicted trip volumes by mode and destination from the ex-toll model, compared with observed values from Replica data. Panels (a) through (f) show mode choice predictions across different geographic areas, including the CRZ, upper Manhattan, the rest of NYC, and the rest of CSA spanning New York, New Jersey, Connecticut, and Pennsylvania. Across all regions, the predicted trip volumes (orange bars) closely match the observed values (blue bars) for each mode. Panel (g) illustrates county-level destination choice predictions, where the model successfully replicates the distribution of trips across destination counties, including the distinction between the CRZ (36061-1) and upper Manhattan (36061-0).

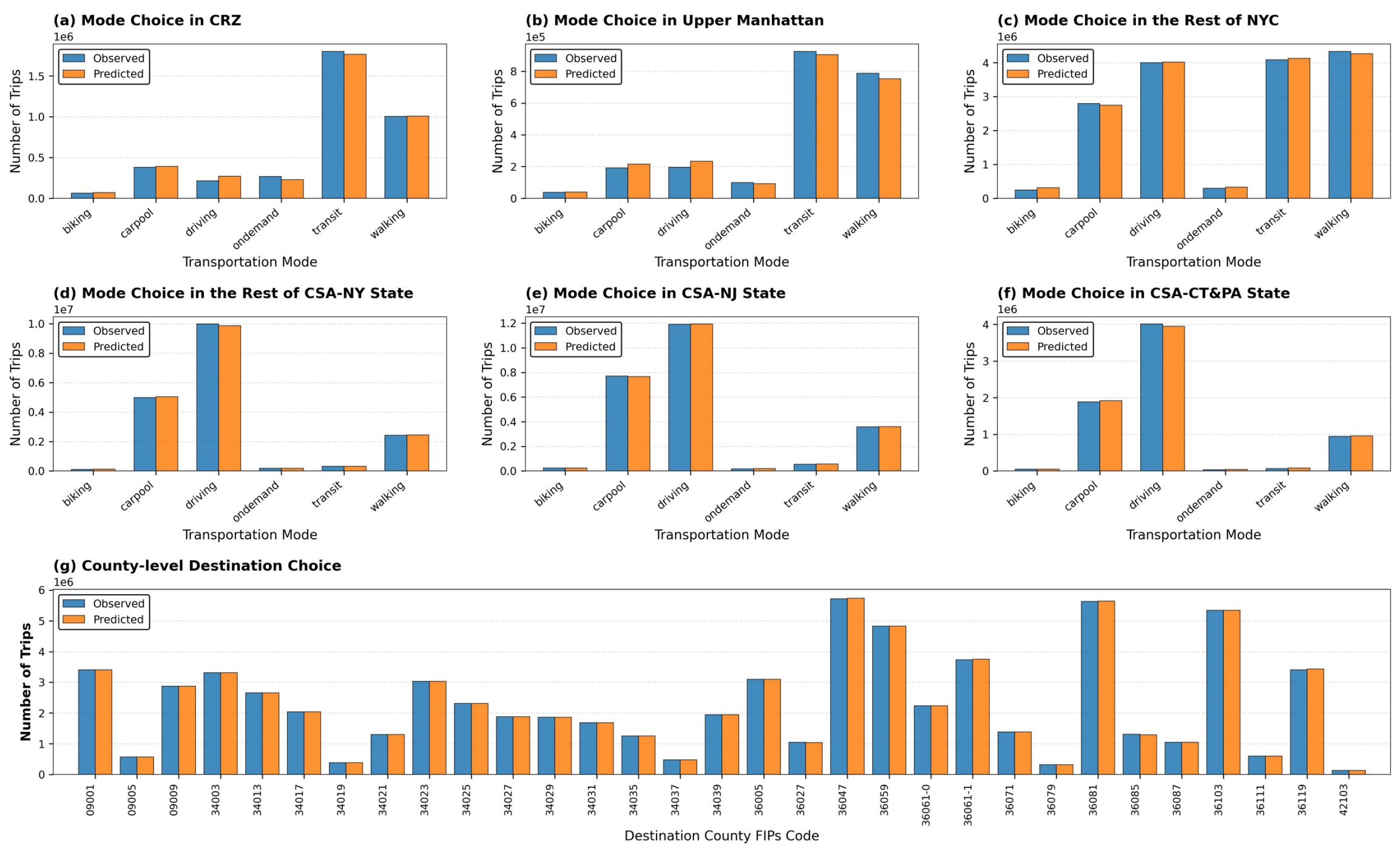

**Fig. 3.** Observed and predicted trip volume by mode and destination. In panel (g), "36061-1" refers to the CRZ and "36061-0" refers to the upper Manhattan.

The toll-related parameters in the post-toll model are calibrated as follows: $\theta_{asc-toll}^{driving} = -0.281$, $\theta_{asc-toll}^{transit} = +0.326$, $\theta_{asc-toll}^{fhv} = +0.170$, $\theta_{asc-toll}^{carpool} = -0.068$, and $\theta_{asc-toll}^{CRZ} = -0.289$. With these calibrated parameters, the post-toll model predictions exactly replicate the observed changes reported in Table 2: a 12.04% reduction in vehicle entries to the CRZ, a 1.41% increase in FHV and taxi trips, and a 9.43% increase in total transit ridership.

The calibrated preference shifts may reflect a range of behavioral mechanisms that emerge after congestion pricing implementation. Travelers who experience reduced street congestion and shorter travel times may revise their expectations and perceive the toll as less burdensome than anticipated. Transit may become relatively more attractive as riders gain firsthand experience with improved service conditions under the post-toll regime. Additionally, travelers who switch destinations in response to the toll may develop new travel habits, treating the CRZ as less attractive. These patterns are consistent with evidence from other congestion pricing implementations (Börjesson et al., 2016). Our calibration framework captures these revealed preference changes in aggregate through the toll-related preference shifts ($\Delta^{toll}\theta_{g,j}$), though it cannot disentangle the specific mechanisms.

Fig. 4 presents the percentage change in trips by mode and destination region predicted by the ex-toll and post-toll models. Regarding mode choice, transit ridership increases substantially across all comparisons, primarily drawing trips from driving, biking, and walking. Although the percentage decrease in driving trips is small, driving accounts for the largest trip volume, so the absolute reduction in driving trips is considerable. Trips destined for the CRZ decrease by 2.73% overall, suggesting that some travelers choose to avoid the congestion pricing zone altogether by redirecting to alternative destinations. These two patterns reflect the intended behavioral responses to congestion pricing: encouraging mode shift toward transit and reducing entries into the charged zone.

Disaggregating by trip segment reveals notable heterogeneity in behavioral responses. Commute trips are generally less sensitive to congestion pricing compared to non-commute trips, consistent with the higher schedule rigidity and lower substitutability of work-related travel. Peak-period trips exhibit larger mode and destination shifts than overnight trips, likely because the daytime congestion toll is substantially higher than the overnight rate. Across population groups, low-income and student travelers show greater sensitivity to the toll: they are more likely to switch to transit and to redirect trips to destinations outside the CRZ.

While the decrease in biking and walking trips may appear counterintuitive given the expected street-level improvements from reduced vehicle traffic, this pattern likely reflects the combined effect of increased transit service attractiveness and the calibrated preference shifts that collectively make transit more competitive relative to active modes in the CRZ. As more granular post-toll data on active transportation become available, future calibration could incorporate biking and walking volumes as additional targets to better capture these dynamics.

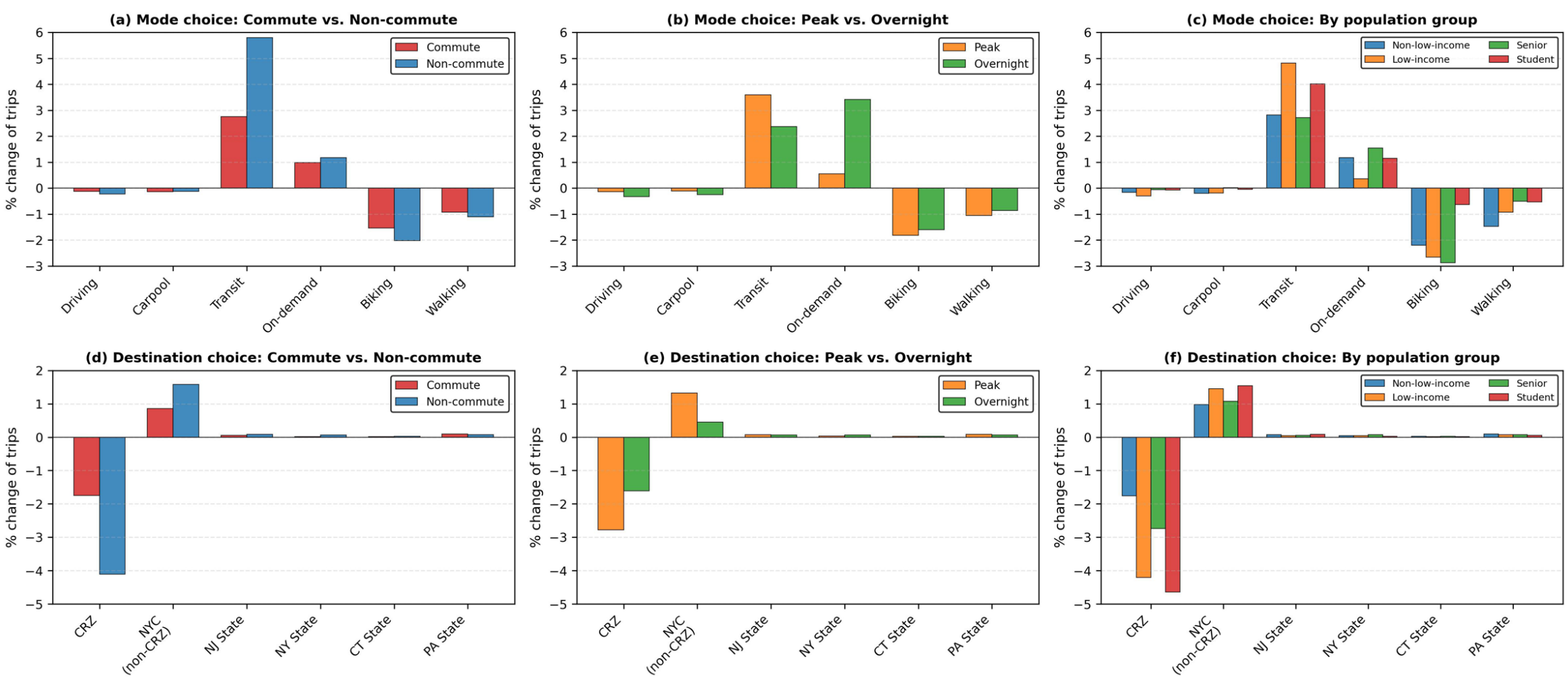

**Fig. 4.** Percentage change in predicted trips by mode and destination disaggregated by trip purpose, time period, and population group. In panels (d)–(f), "CRZ" refers to the Congestion Relief Zone, and "NYC (non-CRZ)" refers to the rest of NYC.

Table 5 compares the gross toll revenue estimated using MTA-reported vehicle entries with estimates derived from post-toll model predictions (MTA, 2026b). To ensure comparability, we focus on two vehicle classes: "Cars, Pickups and Vans," which corresponds to driving and carpool trips in our model (assuming a carpool occupancy of 2, such that two trips account for one vehicle entry), and "TLC Taxi/FHV," which corresponds to on-demand (fhv) trips. We first calculate daily revenue figures for a typical weekday in 2025 Q2, and annual revenue is approximated by multiplying daily revenue by 365 to provide a more interpretable scale.

The model-predicted annual revenue of $941.58 million closely aligns with the MTA-based estimate of $889.14 million, yielding an overall difference of 5.90%. This close correspondence validates the post-toll model's ability to replicate general traffic patterns under the congestion pricing. Notably, both estimates substantially exceed the actual toll revenue of $704.7 million (for all vehicle classes) in the MTA's First Evaluation Report. This discrepancy may be attributable to several factors, including seasonal variation in travel

demand, delays in toll payment processing, and toll exemptions not considered in our calculation.

**Table 5**
Ex-toll model results for selected trip segments (each entry represents the average value, and the number in the parenthesis is the standard error).

| Vehicle Class | Time Period | Toll Rate ($) | Num. Vehicles (per day) | | Annual Toll Revenue (M$) | | |
|---|---|---|---|---|---|---|---|
| | | | MTA | Model | MTA | Model | % Diff |
| Cars, Pickups and Vans | Peak | 9.00 | 226,417 | 248,107 | 743.78 | 815.03 | 9.58% |
| Cars, Pickups and Vans | Overnight | 2.25 | 56,412 | 52,435 | 46.33 | 43.06 | -7.06% |
| TLC Taxi/FHV | Peak | 1.50 | 126,535 | 116,828 | 69.28 | 63.96 | -7.68% |
| TLC Taxi/FHV | Overnight | 1.50 | 54,338 | 35,671 | 29.75 | 19.53 | -34.35% |
| Total | | | | | 889.14 | 941.58 | 5.90% |

### 4.2 Distributional welfare impacts on accessibility

According to the CV decomposition results, the total CV from ex-toll to post-toll period ($CV_t^g$) is $-789,470 per day, in which +$298,840 is due to transit improvement and fuel price reduction and -$1,088,310 (or $397.23 million per year) is due to congestion toll ($CV_t^{toll,g}$). As mentioned in Section 3.1.4, based on the MTA's First Evaluation Report (MTA, 2026a), the estimated annual toll revenue from resident trips is $634.23 million and net revenue is $523.44 million.

Taken together, the total CS loss ($397.23 million per year) accounts for 75.89% of the net toll revenue ($523.44 million per year), demonstrating that the policy satisfies Kaldor–Hicks efficiency (Kaldor, 1939): the net toll revenue alone is sufficient to compensate for the welfare losses experienced by travelers due to the charged zone. This calculation does not account for additional benefits such as environmental improvements from lower emissions and public health gains, suggesting that the true net welfare impact of the policy is likely even more favorable.

However, this does not ensure the policy is Pareto improving, where no one is made worse off by the implementation of a policy (Varian, 1992), as distributional impacts on accessibility persist due to heterogeneities in toll burdens, travel time changes, and how travelers perceive them.

#### *4.2.1 Value of time (VOT) by segment*

Toll burdens and travel time savings are perceived differently across traveler groups, which can be captured using the estimated value of time (VOT). Table 6 lists the VOTs and their standard errors for 16 trip segments. By incorporating NYC-specific interaction terms into travel time and cost, we are able to differentiate VOTs between trips starting from NYC and those from other regions.

Non-low-income commuters during peak hours exhibit the highest VOT across all time components, with auto travel time valued at $39.58 (s.e. = 5.53) per hour in NYC and transit wait time reaching $50.55 (s.e. = 24.46) per hour, underscoring the heightened disutility of delays for this group. By contrast, low-income travelers consistently exhibit lower VOTs—approximately half those of non-low-income travelers for auto travel time ($18.55 vs. $39.58 per hour in NYC during peak commutes)—indicating that they place greater weight on monetary cost relative to travel time savings. Seniors demonstrate moderate VOTs for auto travel time ($12.80 (s.e. = 1.45) per hour for peak commutes in NYC) but show limited sensitivity to transit in-vehicle and wait times, as reflected by large standard errors relative to point estimates (e.g., transit wait time VOT of $17.54 with SE = 14.32), suggesting greater flexibility in scheduling. Students report the lowest VOTs overall, particularly for auto travel time ($9.26 (s.e. = 1.40) per hour for peak commutes in NYC). The NYC-specific interactions further highlight spatial variation: VOTs in NYC are consistently 1.5 to 2 times higher than in other regions across all demographic groups, emphasizing the premium that NYC travelers place on time savings.

**Table 6**

VOT ($/hour) by trip segment and region (each entry represents the point estimation, and the number in the parenthesis is the standard error)

| | VOT (autoTT) | | VOT (transitIVT) | | VOT (transitWT) | | VOT (nonautoTT) | |
|---|---|---|---|---|---|---|---|---|
| | NYC | Other | NYC | Other | NYC | Other | NYC | Other |
| **NotLowIncome Population** | | | | | | | | |
| Commute, Peak | 39.58 | 23.07 | 33.95 | 17.88 | 50.55 | 30.84 | 22.18 | 13.51 |
| | (5.53) | (2.46) | (5.31) | (2.41) | (24.46) | (15.92) | (3.64) | (1.94) |
| Commute, Overnight | 22.04 | 13.67 | 11.18 | 5.38 | -1.05 | 6.93 | 8.98 | 6.45 |
| | (2.56) | (1.45) | (1.66) | (0.96) | (8.03) | (6.21) | (1.47) | (1.06) |
| Non-commute, Peak | 29.66 | 15.28 | 17.80 | 13.40 | 30.87 | 23.82 | 15.46 | 11.14 |
| | (3.45) | (2.45) | (41.75) | (34.99) | (18.24) | (15.31) | (8.65) | (7.27) |
| Non-commute, Overnight | 13.15 | 6.13 | 10.63 | 5.18 | 13.12 | 8.44 | 10.61 | 3.75 |
| | (1.52) | (1.17) | (7.15) | (6.36) | (13.07) | (10.63) | (1.16) | (0.73) |
| **LowIncome Population** | | | | | | | | |
| Commute, Peak | 18.55 | 12.75 | 7.41 | 4.28 | 22.47 | 21.66 | 6.79 | 5.07 |
| | (3.99) | (2.39) | (2.21) | (1.43) | (58.93) | (49.54) | (1.66) | (1.16) |
| Commute, Overnight | 12.39 | 7.58 | 7.98 | 4.49 | -2.46 | 5.69 | 4.90 | 3.99 |
| | (2.21) | (1.29) | (1.51) | (0.86) | (4.71) | (4.03) | (1.00) | (0.78) |
| Non-commute, Peak | 9.22 | 4.85 | 0.06 | -2.28 | 14.79 | 12.77 | 6.96 | 3.11 |
| | (1.48) | (0.93) | (2.66) | (2.12) | (22.13) | (18.59) | (2.18) | (1.73) |
| Non-commute, Overnight | 9.57 | 4.50 | 5.47 | 1.60 | -12.34 | -3.26 | 7.92 | 4.58 |
| | (1.55) | (0.91) | (2.02) | (1.46) | (11.62) | (9.38) | (1.42) | (0.95) |
| **Senior Population** | | | | | | | | |
| Commute, Peak | 12.80 | 8.45 | 10.59 | 5.80 | 17.54 | 13.56 | 7.32 | 5.17 |
| | (1.45) | (0.93) | (2.07) | (1.40) | (14.32) | (11.18) | (1.17) | (0.86) |
| Commute, Overnight | 9.64 | 5.02 | 9.49 | 3.23 | -12.06 | -4.46 | 6.25 | 4.33 |
| | (1.14) | (0.70) | (1.64) | (1.04) | (11.72) | (9.01) | (1.16) | (0.87) |
| Non-commute, Peak | 10.95 | 4.83 | -0.01 | -3.54 | 18.94 | 7.97 | 5.36 | -3.05 |
| | (1.67) | (0.99) | (14.17) | (10.52) | (36.68) | (25.60) | (4.21) | (3.15) |
| Non-commute, Overnight | 9.33 | 3.06 | -3.00 | -5.85 | 0.18 | 4.88 | 12.24 | 7.87 |
| | (1.46) | (0.84) | (10.80) | (7.80) | (22.30) | (14.92) | (2.35) | (1.60) |

| **Student Population** | | | | | | | | |
|---|---|---|---|---|---|---|---|---|
| Commute, Peak | 9.26 | 5.73 | 7.63 | 3.65 | 14.92 | 12.43 | 5.41 | 4.19 |
| | (1.40) | (0.90) | (2.15) | (1.53) | (16.07) | (13.23) | (0.74) | (0.52) |
| Commute, Overnight | 10.91 | 5.63 | 8.94 | 4.53 | -4.51 | -0.94 | 4.05 | 3.09 |
| | (1.69) | (0.92) | (1.66) | (1.01) | (7.54) | (6.15) | (0.76) | (0.57) |
| Non-commute, Peak | 11.99 | 4.88 | 7.02 | 2.20 | 21.40 | 7.35 | 1.76 | -5.00 |
| | (2.00) | (1.14) | (11.21) | (8.29) | (16.96) | (9.64) | (3.91) | (2.96) |
| Non-commute, Overnight | 12.08 | 4.20 | 12.26 | 3.67 | -2.64 | 3.78 | 8.64 | 4.40 |
| | (1.99) | (1.06) | (8.42) | (5.95) | (9.88) | (5.10) | (1.63) | (1.01) |

Note: "autoTT" refers to auto travel time; "transitIVT" refers to transit in-vehicle time; "transitWT" refers to transit wait time; "nonautoTT" refers to nonauto travel time. Standard error of VOT is calculated using the normal approximation approach (Armstrong et al., 2001).

#### *4.2.2 Welfare impacts across regions and segments*

Fig. 5 highlights the uneven spatial distribution of welfare impacts under congestion pricing. The CRZ shows a welfare gain of around \$70,000 per day, while substantial daily welfare losses are concentrated in Queens (around –\$140,000/day), Kings County/Brooklyn (around –\$110,000/day), and upper Manhattan (around –\$100,000/day), reflecting their higher auto trip volumes into the CRZ. Notable losses also occur in nearby New Jersey counties, particularly Hudson (around –\$100,000/day) and Bergen (around –\$100,000/day). Most outer counties in New York State, Connecticut, and Pennsylvania incur relatively small losses, consistent with their lower exposure to the toll. Some peripheral counties exhibit slight positive compensating variation, likely attributable to travel time savings from reduced congestion (Cook et al., 2025; Eliasson et al., 2009). These results emphasize that welfare losses from congestion pricing are highly concentrated in the NYC boroughs and nearby counties with the strongest travel connections to the CRZ.

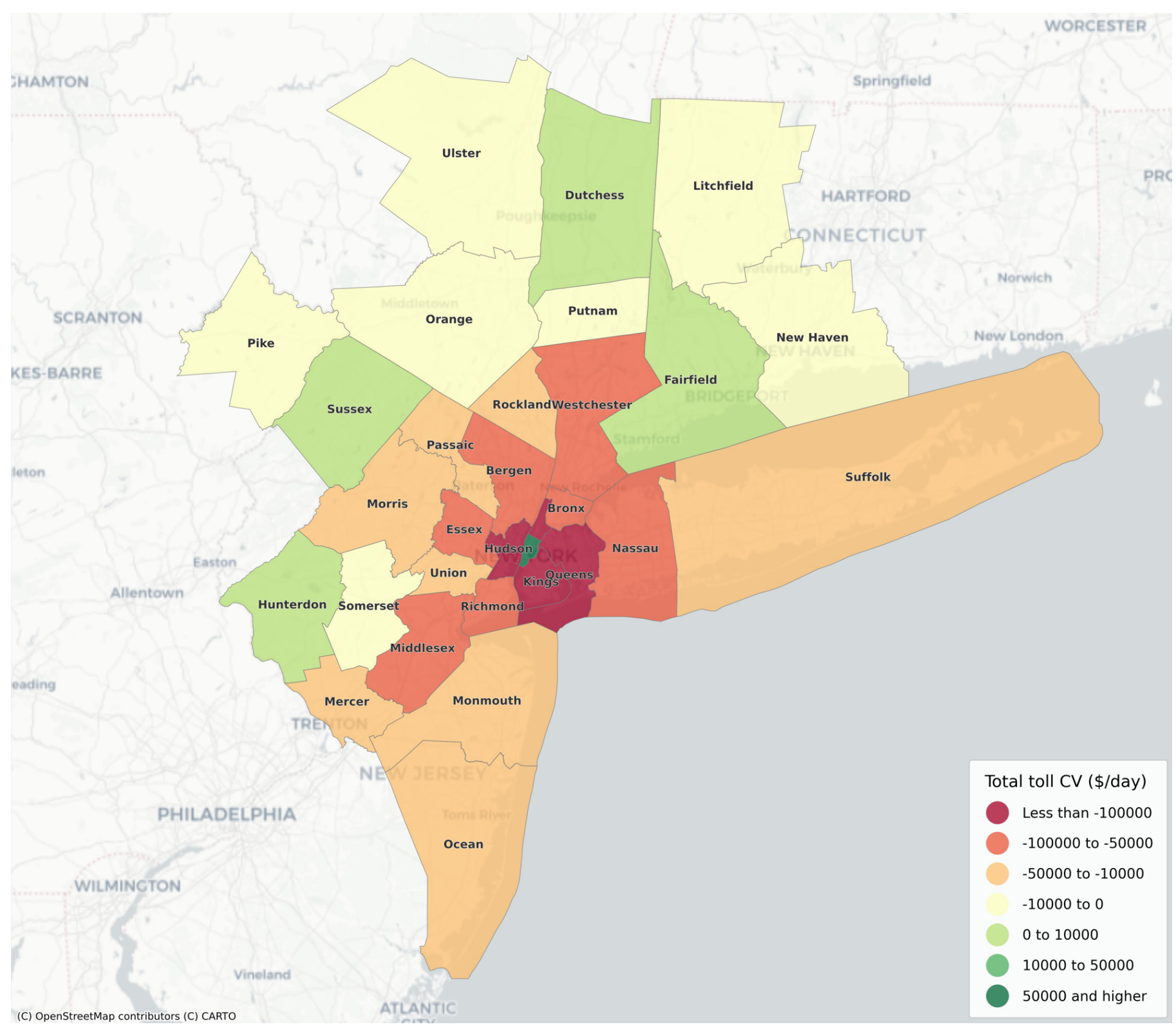

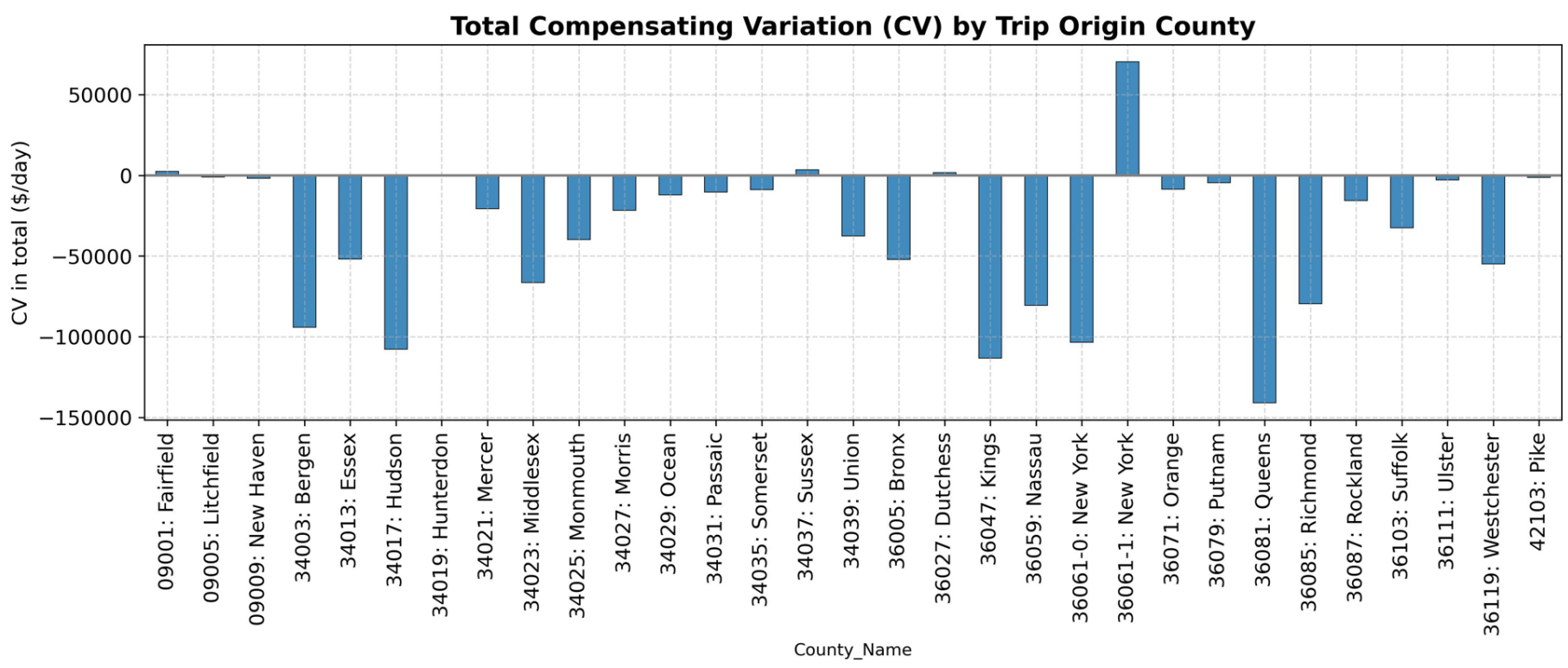

**Fig. 5.** Welfare impacts across trip origin counties. “36061-1” refers to the CRZ and “36061-0” refers to the upper Manhattan.

Fig. 6 presents the average CV ($) for a single trip across different origins, trip purposes, and population groups, where higher values reflect greater welfare impacts at the individual trip level. Commute trips originating from the CRZ exhibit positive CV values, with non-low-income commuters experiencing the largest per-trip gains (around $0.10/trip), while non-commute trips from the CRZ show negative CV values (around –$0.05 to –$0.10/trip). This divergence likely reflects differences in mode dependence and time valuation: CRZ commuters, who have higher values of time and higher transit usage, benefit more from increased transit mode constant and reduced street congestion, whereas non-commute trips from the CRZ are more likely to involve driving or for-hire vehicles, exposing these travelers to the toll costs. Low-income travelers experience smaller welfare losses compared to non-low-income travelers across most origin-purpose combinations, likely because they are more likely to substitute with public transit or redirected trips to alternative destinations.

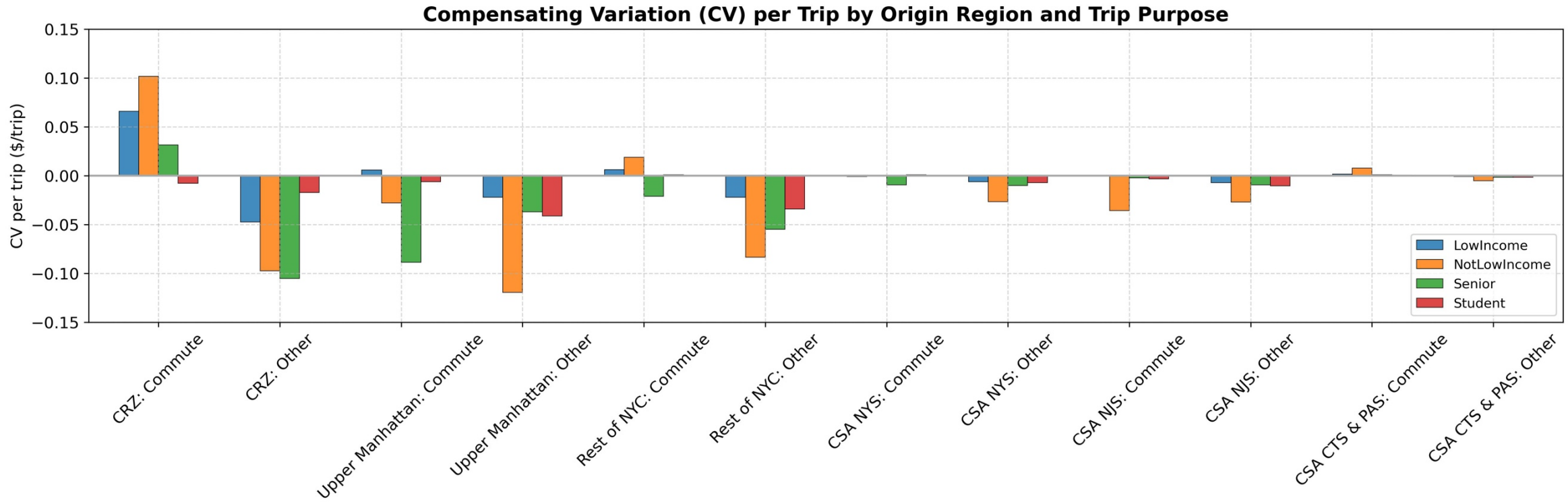

**Fig. 6.** CV ($) per trip aggregated by segments. “Commute” refers to commute trips. “Other” refers to non-commute trips.

### 4.3 Compensatory transit strategies

As the results in Section 4.2 show, welfare losses are concentrated in NYC and New Jersey. Accordingly, we split the set of markets for compensation and consider two subsets: (1) $T_{NYC}$, which includes all markets with trips originating in NYC, where reductions in wait time and fares are applied to all alternatives with transit mode; (2) $T_{NJ}$, which includes all markets with trips originating in New Jersey, where reductions in wait time and fares are applied to all alternatives with transit mode.

*4.3.1 Evaluations under the scenario of Kaldor-Hicks efficiency*

Under the Kaldor–Hicks efficiency scenario, we evaluate two compensatory strategies independently and estimate the extent of transit wait time reduction or fare discount needed to offset the aggregate accessibility-related welfare loss. As shown in Table 7, achieving full welfare compensation requires a reduction of about 0.63 minutes in waiting time for NYC and 2.12 minutes in waiting time for New Jersey. Given the current average wait times (5.00 minutes in NYC and 7.69 minutes in New Jersey), these represent 13% and 28% reductions, respectively.

Alternatively, welfare compensation can be achieved through segment-specific fare discounts, requiring an annual subsidy of approximately $165.15 million for NYC and $171.42 million for New Jersey. The non-low-income population requires the largest fare discount per trip, as their sensitivity to monetary costs is relatively lower than time savings. In contrast, low-income and student travelers, who are more cost-sensitive, can be compensated with smaller reductions in fares. This imbalance raises an equity concern—compensating solely through fare discounts may inadvertently allocate a larger share of financial benefits to higher-income travelers, who are less burdened by monetary costs.

**Table 7**
Compensatory transit strategies under Kaldor-Hicks efficiency

| | Wait time reduction (min) | Segment-level fare discount ($/trip) | Subsidy for discount (M $/year) |
|---|---|---|---|
| NYC part | 0.63 | NotLowIncome:0.3; LowIncome:0.07 Senior:0.44; Student:0.06 | 165.15 |
| New Jersey part | 2.12 | NotLowIncome:2.09; LowIncome:0.31 Senior:1.40; Student:0.41 | 171.42 |

*4.3.2 Evaluations under the Pareto improvement scenario*

Fig. 7 presents combinations of wait time reduction and fare subsidy required to achieve Pareto improvement, along with the corresponding predicted increase in transit ridership after compensation. When no reduction in transit wait time is implemented, the required fare subsidy is approximately $892.76 million per year for NYC and $372.90 million per year for New Jersey, far exceeding the amounts needed under Kaldor–Hicks efficiency. Even with a 100% reduction in transit wait time, a small fare subsidy remains necessary to fully offset welfare losses in New Jersey, highlighting the inefficiency of relying on a single compensatory strategy.

For NYC residents (panel a), the required fare subsidy declines steeply with initial wait time reductions: a 10% decrease in wait time reduces the required subsidy from $892.76 million to approximately $400 million. This suggests that Pareto improvement can be achieved through a combined strategy of modest service improvements and manageable fare subsidies. New Jersey residents (panel b) are less responsive to wait time reduction. An 80% reduction in wait time still requires approximately $50 million in annual fare subsidies. This lower responsiveness may reflect the fact that New Jersey trips tend to be longer and more expensive, making waiting time a smaller fraction of the generalized trip cost, and that OTP-based wait time overstate actual value for lower-frequency New Jersey transit services

where travelers are more likely to consult schedules (see Section 5.1). In addition, the transit ridership response is slightly more elastic in New Jersey compared to NYC, likely because cross-state commuters have greater potential for transit usage when service quality improves.

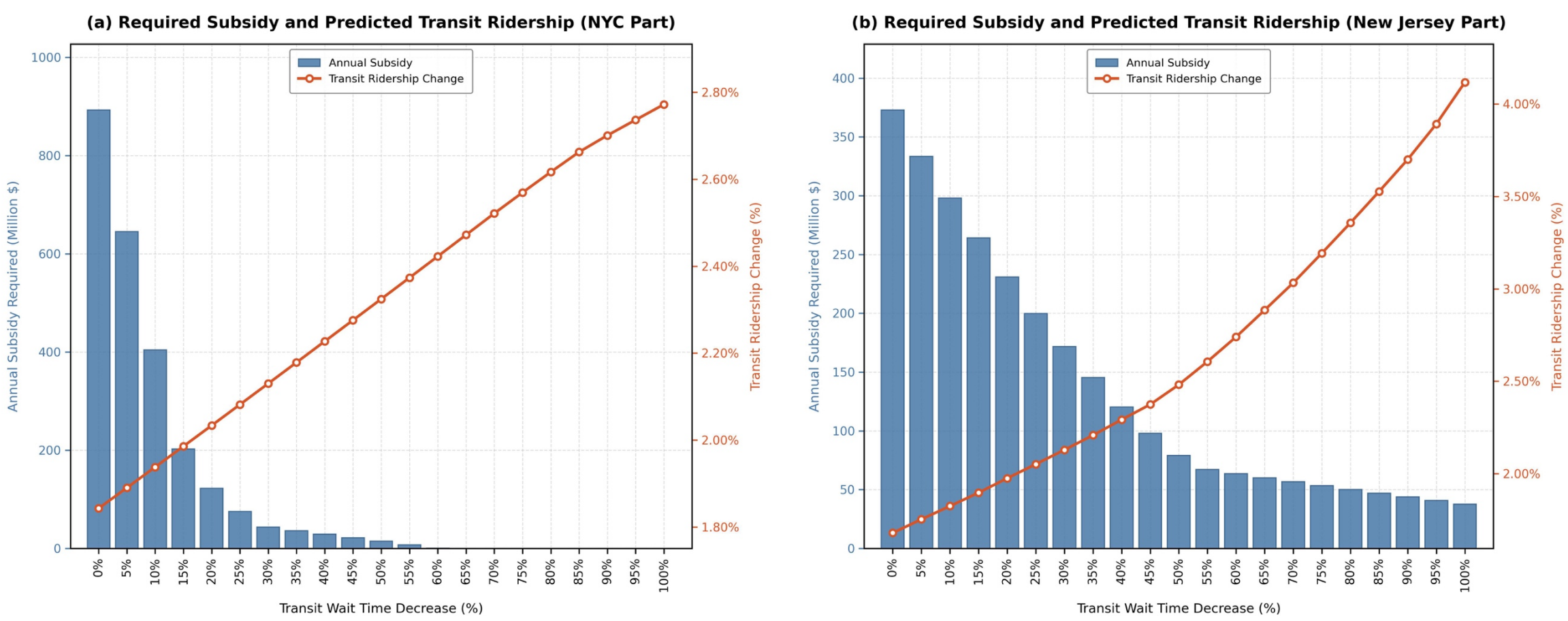

**Fig. 7.** Compensation strategy evaluation under Pareto improvement

Table 8 summarizes the required segment-level fare discounts and corresponding annual subsidies under various levels of wait time reduction from 0-50% for both NYC and New Jersey travelers. Evaluations under the 55–100% time reduction are summarized in Appendix Table A12. With the decrease of wait time, the non-low-income population can be fully compensated, as their higher value of time makes them more responsive to service frequency improvements. In contrast, senior population and students in New Jersey exhibit lower time sensitivity, still requiring fare discounts even under notable wait time reductions. Moreover, the fare discounts required to achieve Pareto improvement are considerably higher than those needed for Kaldor–Hicks efficiency, especially when the wait time reduction is less than 15%. This indicates that applying uniform discounts leads to overcompensation for some groups, thereby inflating the total subsidy. These results highlight the inefficiency of uniform fare discounts and underscore the importance of differentiated, region-specific adjustments to improve welfare without imposing excessive fiscal burden.

**Table 8**

Compensatory transit strategies under Pareto improvement (0–50% wait time reduction)

| Wait time reduction (%) | **NYC part** | | **New Jersey part** | |
|---|---|---|---|---|
| | Segment-level fare discount ($/trip) | Subsidy for discount (M $/year) | Segment-level fare discount ($/trip) | Subsidy for discount (M $/year) |
| 0 | NotLowIncome: 1.07<br>LowIncome: 0.33<br>Senior: 0.98<br>Student: 0.36 | 892.76 | NotLowIncome: 2.55<br>LowIncome: 1.04<br>Senior: 1.73<br>Student: 0.83 | 372.90 |
| 5 | NotLowIncome: 0.76<br>LowIncome: 0.19<br>Senior: 0.84<br>Student: 0.26 | 645.14 | NotLowIncome: 2.30<br>LowIncome: 0.56<br>Senior: 1.61<br>Student: 0.80 | 333.56 |
| 10 | NotLowIncome: 0.45<br>LowIncome: 0.07 | 404.00 | NotLowIncome: 2.06<br>LowIncome: 0.36 | 297.90 |
| | Senior: 0.69<br>Student: 0.15 | | Senior: 1.49<br>Student: 0.77 | |
| 15 | NotLowIncome: 0.18<br>LowIncome: 0.04<br>Senior: 0.55<br>Student: 0.05 | 202.57 | NotLowIncome: 1.82<br>LowIncome: 0.32<br>Senior: 1.38<br>Student: 0.75 | 264.09 |
| 20 | NotLowIncome: 0.10<br>LowIncome: 0.01<br>Senior: 0.40<br>Student: 0 | 122.48 | NotLowIncome: 1.58<br>LowIncome: 0.32<br>Senior: 1.26<br>Student: 0.74 | 230.91 |
| 25 | NotLowIncome: 0.04<br>LowIncome: 0<br>Senior: 0.34<br>Student: 0 | 75.23 | NotLowIncome: 1.35<br>LowIncome: 0.32<br>Senior: 1.15<br>Student: 0.73 | 199.80 |
| 30 | NotLowIncome: 0<br>LowIncome: 0<br>Senior: 0.29<br>Student: 0 | 43.53 | NotLowIncome: 1.14<br>LowIncome: 0.28<br>Senior: 1.04<br>Student: 0.72 | 171.78 |
| 35 | NotLowIncome: 0<br>LowIncome: 0<br>Senior: 0.24<br>Student: 0 | 36.35 | NotLowIncome: 0.95<br>LowIncome: 0.28<br>Senior: 0.95<br>Student: 0.71 | 145.33 |
| 40 | NotLowIncome: 0<br>LowIncome: 0<br>Senior: 0.20<br>Student: 0 | 29.18 | NotLowIncome: 0.76<br>LowIncome: 0.28<br>Senior: 0.86<br>Student: 0.71 | 120.29 |
| 45 | NotLowIncome: 0<br>LowIncome: 0<br>Senior: 0.15<br>Student: 0 | 22.01 | NotLowIncome: 0.60<br>LowIncome: 0.28<br>Senior: 0.81<br>Student: 0.70 | 97.94 |
| 50 | NotLowIncome: 0<br>LowIncome: 0<br>Senior: 0.10<br>Student: 0 | 14.83 | NotLowIncome: 0.46<br>LowIncome: 0.28<br>Senior: 0.75<br>Student: 0.70 | 79.20 |

## 5. Discussion

This section discusses the key assumptions underlying our modeling framework and policy implications for the design and governance of congestion pricing compensation.

### 5.1 Assumptions and limitations of the modeling framework

Our analysis rests on several simplifying assumptions that merit explicit acknowledgment. First, we aggregate trips to the county level and assume homogeneous tastes within each segment defined by population group, trip purpose, and time period. County-level aggregation aligns with the administrative boundaries at which equity concerns are typically raised and policies are implemented, and the 16-segment stratification (by income, age, trip purpose, and time-of-day) captures the most policy-relevant dimensions of taste variation. Nevertheless, this market-level logit formulation suppresses intra-market heterogeneity that finer-grained data would reveal. For example, residents within the same county may differ substantially in their proximity to transit stations, income levels, and age profiles, leading to heterogeneous experiences of the toll burden that our model cannot distinguish. Sub-county equity assessments would require finer spatial resolution (e.g., census tract or neighborhood level), which we leave for future work as individual-level validation of synthetic trips becomes feasible. In addition, introducing additional region-specific interactions (e.g., for upstate New York, New Jersey, or Connecticut) is feasible within our model framework but would substantially increase the number of taste parameters; we leave this extension for future work with nonparametric estimation approaches such as GLAM logit (Ren et al., 2025).

Second, we rely on synthetic trip data generated by Replica Inc. rather than on observed household travel surveys. Although synthetic populations are calibrated against census, mobile-phone, and economic-activity data, individual-level validation remains limited. At the aggregate level, however, our cross-validation against NYMTC entry counts demonstrates that the synthetic data reproduce observed travel patterns with reasonable fidelity, lending confidence to market-level inference.

Third, the post-implementation model is calibrated using a minimum-norm pseudoinverse approach to match three observed marginal changes (vehicle entries, on-demand trips, and transit ridership) with five toll-related parameters. We seek to use consistent and authoritative values from the MTA's First Evaluation Report. Introducing additional calibration targets may improve the robustness but would require data from sources with potentially different collection methodologies or spatial boundaries, which could also introduce inconsistencies into the calibration. The system is thus underdetermined, and the pseudoinverse selects the smallest parameter adjustment consistent with the targets. This parsimony principle is a defensible first-order strategy, but it cannot capture richer behavioral shifts (e.g., effects on travelers' sensitivity to travel time and cost) that may emerge as the program matures. Moreover, although the toll-related preference shifts ($\Delta^{toll}\theta_{g,j}$) are calibrated to match observed changes in the Congestion Pricing First Evaluation Report, these calibrated constants may partly absorb concurrent influences such as income shifts, inflation, or evolving remote work patterns that are not explicitly controlled for in our model. In addition, our model treats trips to destinations outside the CSA as the outside alternative with a fixed market-specific constant. In practice, congestion pricing may also affect the attractiveness of these outside destinations, which our current specification does not capture.

Fourth, the waiting time used in our model is computed by OpenTripPlanner as one-half the scheduled headway, which is an approximation given that no publicly available dataset provides observed waiting times across our study area. This approximation may overestimate actual waiting time for low-frequency routes where travelers consult schedules and time their arrivals, and may understate waiting time on routes with unreliable service. More broadly, while our study use wait time reduction as a main proxy for transit improvement, transit agencies have other operational levers at their disposal, such as route restructuring to reduce transfers, dedicated bus lanes to increase vehicle speeds (reduce in-vehicle travel time), and service reliability enhancements that reduce headway variability. Future work should evaluate their individual effects and corresponding cost-benefit tradeoffs.

### 5.2 Policy implications for congestion toll compensation

Our results carry several interconnected implications for the design of compensatory strategies accompanying congestion tolls. A first insight is that no single compensatory lever is sufficient. Segment-specific values of time differ widely: non-low-income peak commuters in NYC place a high premium on transit wait time, whereas low-income and student travelers are more sensitive to monetary cost. Consequently, frequency improvements alone leave cost-sensitive groups under-compensated, while fare discounts alone are expensive and poorly targeted for high-VOT segments. The efficient frontier therefore combines both instruments: modest, broadly distributed headway reductions that benefit all travelers (especially high-VOT users) alongside targeted, means-tested fare relief for groups whose primary barrier is cost. We note that the NYC metropolitan area already employs distance-based fare structures for certain services, such as commuter rail (Metro-North and LIRR) and the PATH system. These existing distance-based structures could serve as a practical framework for implementing more spatially targeted fare relief in future policy design.

A second insight concerns the choice of compensation criterion. Achieving Kaldor–Hicks efficiency requires substantially fewer resources than pursuing Pareto improvement for every population group and county, but the former can mask significant distributional shortfalls. Two implications follow. First, when the policy target is aggregate efficiency (e.g., to ensure the program's fiscal viability), agencies should still track a small set of “equity sentinel” segments, such as auto-reliant New Jersey residents and specific NYC neighborhoods with limited alternatives. Second, where a jurisdiction aims for stronger equity guarantees, Pareto-style goals should be operationalized through more granular instruments (e.g., origin-specific discounts or commuter-product bundles) rather than uniform fare reductions, to avoid over-spending on groups already fully compensated. However, finer-grained targeting also raises administrative costs and opens the door to political lobbying for exemptions and carveouts that can erode the program's core benefits (Komanoff, 2019). In practice, agencies should therefore balance equity granularity against administrative complexity and political vulnerability, and can stage compensation accordingly: first meet a Kaldor–Hicks threshold system-wide, then add focused, data-driven transfers until identified gaps close, while resisting pressure to broaden exemptions beyond evidence-based thresholds.

Third, the compensation portfolio should extend beyond transit. Trucks and commercial vehicles pay higher tolls yet cannot substitute to bus or subway. This highlights the need for a broader compensation portfolio that complements transit with freight-oriented measures. Examples include delivery consolidation support, off-peak delivery incentives (paired with curb management and enforcement), and grants for zero-emission freight vehicles that both reduce operating costs and amplify air-quality benefits near the cordon. Similarly, some auto-dependent neighborhoods may require “first/last-mile” connectors (e-bike share, microtransit) to make transit frequency gains usable. Finally, environmental and health co-benefits—documented in other cities and emerging in New York—should be captured explicitly in the reinvestment calculus: bus priority and signal priority in the CRZ can lock in speed gains; targeted station accessibility upgrades ensure benefits accrue to seniors and people with disabilities; and sidewalk/bike-network investments expand viable non-auto substitutes. These complementary measures do not replace frequency and fare tools; they make them effective for populations whose constraints lie outside the transit farebox. However, this would bring potential increases of administrative costs that are not captured by this study.

While our analysis focuses on compensatory strategies to offset welfare losses, it is important to recognize that congestion pricing generates net welfare gains even before any reinvestment occurs. Welfare would increase even if the revenue were unused. However, reinvestment is essential to ensure that no traveler group is made worse off and to maintain public acceptance of the program. Moreover, investing toll revenue in transit improvements may generate welfare gains beyond simple compensation, given the marginal cost of public funds: each dollar of toll revenue avoids the economic distortions associated with raising revenue through taxes or other sources, effectively providing a premium on public investment. In practice, NYC's congestion pricing revenue is legally mandated to fund capital improvements to public transit and commuter rail service, aligning fiscal constraints with the compensatory strategies evaluated in this study.

## 6. Conclusion

NYC's congestion pricing program offers a rare, real-world testbed for how congestion pricing reshapes travel behavior, welfare, and policy priorities in a dense metropolitan region. Motivated by persistent concerns over fairness and practicality, this study measures distributional welfare impacts across New York and New Jersey and evaluates transit reinvestment strategies that can credibly compensate losses. To that end, we leverage a regional, joint mode and destination framework that connects ex- and post-toll conditions, allowing us to speak directly to questions that matter for program legitimacy: who bears losses, where, and by how much—plus what mix of transit frequency improvements and fare relief most effectively restores welfare.

Methodologically, the study advances welfare analysis with joint mode and destination choice models estimated using large-scale synthetic trips. Parameters reflecting toll-related preference changes are calibrated using numerical values from the authoritative report and validated against MTA vehicle entries. Substantively, results show small changes in overall trip-making but clear mode and destination shifts surrounding the CRZ. Net welfare is positive once toll revenue is included, yet losses are unequally distributed—concentrated in upper Manhattan, Brooklyn, Queens, and Hudson County. Value-of-time heterogeneity is pronounced: high-VOT commuters are especially sensitive to waiting, while cost-sensitive groups respond more to fares. Consequently, single-lever compensation performs poorly. A mixed strategy—modest, broadly applied wait-time reductions paired with targeted fare discounts—can achieve Pareto improvement at manageable fiscal cost. Pareto-style guarantees become much more expensive and often infeasible via service improvements alone.

Despite simplifications and limitations discussed in Section 5.1, we contend that the framework is reasonable for a first-order welfare analysis of congestion pricing. The logit structure yields closed-form consumer-surplus expressions that are transparent and reproducible; segment-specific parameters preserve the taste heterogeneity most relevant to equity evaluation; and the calibration targets are drawn from the best available post-toll evidence. These choices strike a deliberate balance between methodological rigor and practical tractability, and they provide a structured foundation on which future studies with richer post-rollout panels and individual-level data can build.

## References

American Trucking Associations. (2025). *Trucking Renews Push to End NYC Congestion Pricing Ahead of March 21 Deadline*. https://www.trucking.org/news-insights/trucking-renews-push-end-nyc-congestion-pricing-ahead-march-21-deadline. Retrieved February 7, 2026

Angrist, J. D., & Krueger, A. B. (2001). Instrumental Variables and the Search for Identification: From Supply and Demand to Natural Experiments. *Journal of Economic Perspectives*, *15*(4), 69–85.

Armstrong, P., Garrido, R., & de Dios Ortúzar, J. (2001). Confidence intervals to bound the value of time. *Transportation Research Part E: Logistics and Transportation Review*, *37*(2–3), 143–161.

Bachir, D., Khodabandelou, G., Gauthier, V., El Yacoubi, M., & Puchinger, J. (2019). Inferring dynamic origin-destination flows by transport mode using mobile phone data. *Transportation Research Part C: Emerging Technologies*, *101*, 254–275.

Baghestani, A., Tayarani, M., Allahviranloo, M., Nadafianshahamabadi, R., Kucheva, Y., Reza Mamdoohi, A., & Oliver Gao, H. (2022). New York City cordon pricing and its impacts on disparity, transit accessibility, air quality, and health. *Case Studies on Transport Policy*, *10*(1), 485–499.

Basso, L. J., & Jara-Díaz, S. R. (2012). Integrating congestion pricing, transit subsidies and mode choice. *Transportation Research Part A: Policy and Practice*, *46*(6), 890–900.

Berry, S., Levinsohn, J., & Pakes, A. (1995). Automobile prices in market equilibrium. *Econometrica: Journal of the Econometric Society*, 841–890.

Berry, S. T. (1994). Estimating Discrete-Choice Models of Product Differentiation. *The RAND Journal of Economics*, *25*(2), 242.

Bills, T. S., Twumasi-Boakye, R., Broaddus, A., & Fishelson, J. (2022). Towards transit equity in Detroit: An assessment of microtransit and its impact on employment accessibility. *Transportation Research Part D: Transport and Environment*, *109*, 103341.

Bloomberg, M. (2007). *PlaNYC: A Greener, Greater New York*. https://www.nyc.gov/html/planyc/downloads/pdf/publications/full_report_2007.pdf

Börjesson, M., Eliasson, J., & Hamilton, C. (2016). Why experience changes attitudes to congestion pricing: The case of Gothenburg. *Transportation Research Part A: Policy and Practice*, *85*, 1–16.

Bowman, J. L., & Ben-Akiva, M. E. (2001). Activity-based disaggregate travel demand model system with activity schedules. *Transportation Research Part A: Policy and Practice*, *35*(1), 1–28.

Chen, C. (2025). *Roots of resistance: Understanding the opposition to congestion pricing in New York City*. http://caee.webhost.utexas.edu/prof/kockelman/public_html/BTR7rootsofresistance.pdf. Retrieved February 7, 2026

Chen, R., & Nozick, L. (2016). Integrating congestion pricing and transit investment planning. *Transportation Research Part A: Policy and Practice*, *89*, 124–139.

Cook, C., Kreidieh, A., Vasserman, S., Allcott, H., Arora, N., van Sambeek, F., Tomkins, A., & Turkel, E. (2025). The Short-Run Effects of Congestion Pricing in New York City. *NBER Working Paper No. 33584*. https://www.nber.org/papers/w33584. Retrieved February 7, 2026

De Palma, A., & Lindsey, R. (2011). Traffic congestion pricing methodologies and technologies. *Transportation Research Part C: Emerging Technologies*, *19*(6), 1377–1399.

Downs, A. (2005). *Still stuck in traffic: Coping with peak-hour traffic congestion*. Rowman & Littlefield.

Eliasson, J. (2009). A cost–benefit analysis of the Stockholm congestion charging system. *Transportation Research Part A: Policy and Practice*, *43*(4), 468–480.

Eliasson, J., Hultkrantz, L., Nerhagen, L., & Rosqvist, L. S. (2009). The Stockholm congestion–charging trial 2006: Overview of effects. *Transportation Research Part A: Policy and Practice*, *43*(3), 240–250.

Fosgerau, M., Julien Monardo, & Palma, A. D. (2024). The inverse product differentiation logit model. *American Economic Journal: Microeconomics*, *16*(4), 329–370.

Freeman III, A. M., Herriges, J. A., & Kling, C. L. (2014). *The measurement of environmental and resource values: Theory and methods*. Routledge.

Ghassabian, A., Titus, A. R., Conderino, S., Azan, A., Weinberger, R., & Thorpe, L. E. (2024). Beyond traffic jam alleviation: Evaluating the health and health equity impacts of New York City's congestion pricing plan. *Journal of Epidemiology and Community Health*, *78*(5), 273–276.

Harris, W., & Ley, A. (2024). How the Resurrected Congestion Pricing Plan Could Die in the Courts. *The New York Times*.

He, B. Y., Zhou, J., Ma, Z., Wang, D., Sha, D., Lee, M., Chow, J. Y., & Ozbay, K. (2021). A validated multi-agent simulation test bed to evaluate congestion pricing policies on population segments by time-of-day in New York City. *Transport Policy*, *101*, 145–161.

Hensher, D. A., & Ho, C. Q. (2016). Experience conditioning in commuter modal choice modelling – Does it make a difference? *Transportation Research Part E: Logistics and Transportation Review*, *95*, 164–176.

Hörl, S., & Balac, M. (2021). Synthetic population and travel demand for Paris and Île-de-France based on open and publicly available data. *Transportation Research Part C: Emerging Technologies*, *130*, 103291.

Huo, J., Dua, R., & Bansal, P. (2024). Inverse product differentiation logit model: Holy grail or not? *Energy Economics*, *131*, 107379.

Hwang, H.-L., Wilson, D. W., Reuscher, T., Chin, S.-M., & Taylor, R. D. (2015). *Travel Patterns and Characteristics of Transit Users in New York State*. Oak Ridge National Lab.(ORNL), Oak Ridge, TN (United States).

Isaksen, E. T., & Johansen, B. G. (2025). Congestion pricing with electric vehicle exemptions: Car-ownership effects and other behavioral adjustments. *Journal of Environmental Economics and Management*, *131*, 103154.

Ji, Y. (2025). Distributional Impacts of Congestion Pricing in New York City. *2025 AAEA & WAEA Joint Annual Meeting*. https://doi.org/https://doi.org/10.22004/ag.econ.360946

Kaldor, N. (1939). Welfare propositions of economics and interpersonal comparisons of utility. *The Economic Journal*, *49*(195), 549–552.

Kockelman, K. M., & Kalmanje, S. (2005). Credit-based congestion pricing: A policy proposal and the public's response. *Transportation Research Part A: Policy and Practice*, *39*(7–9), 671–690.

Komanoff, C. (2019). *Congestion pricing carveouts will steal millions of hours and billions of bucks*. https://nyc.streetsblog.org/2019/03/28/komanoff-congestion-pricing-carveouts-will-steal-millions-of-hours-and-billions-of-bucks, Retrieved May 17, 2026

Krueger, R., Bierlaire, M., & Bansal, P. (2023). A data fusion approach for ride-sourcing demand estimation: A discrete choice model with sampling and endogeneity corrections. *Transportation Research Part C: Emerging Technologies*, *152*, 104180.

Leape, J. (2006). The London congestion charge. *Journal of Economic Perspectives*, *20*(4), 157–176.

Li, W., Kockelman, K. M., & Huang, Y. (2021). Traffic and Welfare Impacts of Credit-Based Congestion Pricing Applications: An Austin Case Study. *Transportation Research Record: Journal of the Transportation Research Board*, *2675*(1), 10–24.

Marazi, N. F., Majumdar, B. B., & Sahu, P. K. (2024). Examining Congestion Pricing Scheme Effectiveness Using the Travel Time Congestion Index. *Transportation Research Record: Journal of the Transportation Research Board*, *2678*(11), 474–488.

McFadden, D. (1972). *Conditional logit analysis of qualitative choice behavior*. https://escholarship.org/uc/item/61s3q2xr

McFadden, D. (1977). Modelling the choice of residential location. *Cowles Foundation Discussion Papers*, *710*.

Meier, A. (2024, November 15). New NYC congestion pricing plan: Details for drivers. *FOX 5 NY*. https://www.fox5ny.com/news/nyc-congestion-pricing-plan-new-tolls-details-drivers-hochul. Retrieved February 7, 2026

Metropolitan Transportation Authority. (2025a). *Central Business District Tolling Program*. https://www.mta.info/project/CBDTP, Retrieved February 7, 2026

Metropolitan Transportation Authority. (2025b). *MTA Releases Revenue From Congestion Relief Zone Tolling Showing Program in Line With Projections*. https://www.mta.info/press-release/mta-releases-revenue-congestion-relief-zone-tolling-showing-program-line-projections. Retrieved February 7, 2026

Metropolitan Transportation Authority. (2026a). *Congestion Relief Zone Tolling First Evaluation Report*. https://www.mta.info/document/195631, Retrieved February 7, 2026

Metropolitan Transportation Authority. (2026b). *Vehicle entries to the Congestion Relief Zone (CRZ) as detected by the tolling system*. https://metrics.mta.info/?cbdtp/vehicleentries. Retrieved February 7, 2026

Mobility Database. (2026). *Global Transit Data* [Data set]. https://mobilitydatabase.org/. Retrieved January 21, 2026

Moore, E. H. (1920). On the reciprocal of the general algebraic matrix. *Bulletin of the American Mathematical Society*, *26*, 294–295.

National Bureau of Economic Research. (2025). Impact of New York City's Congestion Pricing Program. *NBER Digest (June 2025)*. https://www.nber.org/digest/202506/impact-new-york-citys-congestion-pricing-program. Retrieved February 7, 2026

New York Metropolitan Transportation Council. (2026). *Hub Bound Travel*. https://www.nymtc.org/Data-and-Modeling/Transportation-Data-and-Statistics/Publications/Hub-Bound-Travel. Retrieved February 8, 2026

Newman, J. P., & Bernardin Jr, V. L. (2010). Hierarchical ordering of nests in a joint mode and destination choice model. *Transportation*, *37*(4), 677–688.

Niemeier, D. A. (1997). Accessibility: An evaluation using consumer welfare. *Transportation*, *24*(4), 377–396.

Nogueira, M. (2025). The Impact of New York's 2025 Congestion Pricing on Traffic Volume. *Available at SSRN 5383652*.

Osuna, E. E., & Newell, G. F. (1972). Control strategies for an idealized public transportation system. *Transportation Science*, *6*(1), 52–72.

Parr, S., Wolshon, B., Renne, J., Murray-Tuite, P., & Kim, K. (2020). Traffic impacts of the COVID-19 pandemic: Statewide analysis of social separation and activity restriction. *Natural Hazards Review*, *21*(3), 4020025.

Paulley, N., Balcombe, R., Mackett, R., Titheridge, H., Preston, J., Wardman, M., Shires, J., & White, P. (2006). The demand for public transport: The effects of fares, quality of service, income and car ownership. *Transport Policy*, *13*(4), 295–306.

Ren, X., & Chow, J. Y. J. (2022). A random-utility-consistent machine learning method to estimate agents' joint activity scheduling choice from a ubiquitous data set. *Transportation Research Part B: Methodological*, *166*, 396–418.

Ren, X., Chow, J. Y. J., & Bansal, P. (2025). Nonparametric mixed logit model with market-level parameters estimated from market share data. *Transportation Research Part B: Methodological*, *196*, 103220.

Ren, X., Chow, J. Y. J., & Guan, C. (2024). Mobility service design with equity-aware choice-based decision-support tool: New York case study. *Transportation Research Part D: Transport and Environment*, *132*, 104255.

Replica Inc. (2024). *Seasonal Mobility Model Methodology Extended (Places)*. https://documentation.replicahq.com/docs/seasonal-mobility-model-methodology-extended-places

Salon, D. (2009). Neighborhoods, cars, and commuting in New York City: A discrete choice approach. *Transportation Research Part A: Policy and Practice*, *43*(2), 180–196.

Santos, G., & Bhakar, J. (2006). The impact of the London congestion charging scheme on the generalised cost of car commuters to the city of London from a value of travel time savings perspective. *Transport Policy*, *13*(1), 22–33.

Schaller, B. (2010). New York City's congestion pricing experience and implications for road pricing acceptance in the United States. *Transport Policy*, *17*(4), 266–273.

Schwartz, S., Soffian, G., Kim, J. M., & Weinstock, A. (2008). A Comprehensive Transportation Policy for the 21st Century: A Case Study of Congestion Pricing in New York City. *NYU Envtl. LJ*, *17*, 580.

Shalma, M. (2025). Congestion pricing threatens small businesses, events and communities. *Op-Ed*. https://www.amny.com/oped/op-ed-congestion-pricing-threatens-small-business/

Shapley, L. S. (1953). *A value for n-person games*.

Shorrocks, A. F. (2013). Decomposition procedures for distributional analysis: A unified framework based on the Shapley value. *Journal of Economic Inequality*, *11*(1), 99–126.

Siena College Research Institute. (2024). *New York State poll: New Yorkers overwhelmingly oppose congestion pricing plan*. https://sri.siena.edu

Simeonova, E., Currie, J., Nilsson, P., & Walker, R. (2021). Congestion pricing, air pollution, and children's health. *Journal of Human Resources*, *56*(4), 971–996.

Small, K. A. (2012). Valuation of travel time. *Economics of Transportation*, *1*(1–2), 2–14.

Small, K. A., & Rosen, H. S. (1981). Applied welfare economics with discrete choice models. *Econometrica: Journal of the Econometric Society*, 105–130.

Standen, C., Greaves, S., Collins, A. T., Crane, M., & Rissel, C. (2019). The value of slow travel: Economic appraisal of cycling projects using the logsum measure of consumer surplus. *Transportation Research Part A: Policy and Practice*, *123*, 255–268.

Tarduno, M. (2022). For whom the bridge tolls: Congestion, air pollution, and second-best road pricing. *Unpublished Manuscript*.

Varian, H. R. (1992). *Microeconomic analysis* (Vol. 3). Norton New York.

Vij, A., & Walker, J. L. (2016). *Latent Class Choice Models with Feedback through Consumer Surplus*. Citeseer.

Vovsha, P. (1997). Application of cross-nested logit model to mode choice in Tel Aviv, Israel, metropolitan area. *Transportation Research Record*, *1607*(1), 6–15.

Wikipedia contributors. (n.d.). *New York metropolitan area*. Wikipedia. Retrieved 16 January 2026, from https://en.wikipedia.org/wiki/New_York_metropolitan_area

Wongpiromsarn, T., Xiao, N., You, K., Sim, K., Xie, L., Frazzoli, E., & Rus, D. (2012). Road pricing for spreading peak travel: Modeling and design. *arXiv Preprint arXiv:1208.4589*.

Young, M. (2018). *OpenTripPlanner-creating and querying your own multi-modal route planner*. https://github.com/marcusyoung/otp-tutorial. Retrieved February 7, 2026

## Appendix

### A1. Validation of Replica's data

According to Replica's data quality report for the entire U.S. region[1], the largest error of demographic attributes for a single census tract unit is within 5% compared to census data, and the largest error of commute mode share for a single census tract unit is within 10% compared to CTPP data.

Tables A1 – A2 further compare inbound trip volumes and proportions to the Congestion Relief Zone (CRZ), using data from the New York Metropolitan Transportation Council (NYMTC)[2] and Replica. The year 2023 is selected because it is the year for which Replica's synthetic trip data are available. Table A1 examines inbound travel mode distributions, while Table A2 compares entry sector distributions across major corridors identified by NYMTC.

Overall, the two tables indicate a high degree of consistency between Replica and NYMTC data, with percent differences generally within 10% for most aggregate categories. The total inbound volume estimated by Replica is 6.54% lower than that reported by NYMTC, which is expected given that NYMTC data include truck trips whereas we excluded these trips from Replica data. Differences in non-motorized travel appear to arise primarily from mode classification, with some biking trips in NYMTC likely identified as walking in Replica; however, the combined volume of non-motorized trips remains comparable across the two datasets. These results suggest that Replica provides a reasonably accurate representation of inbound travel patterns to the CRZ in this study context, supporting its use for subsequent analysis of congestion pricing impacts.

**Table A1**

Comparison of inbound travel mode distribution: NYMTC vs. Replica Data

| | **NYMTC Data** | | **Replica Data** | | **Percent Difference** | |
|---|---|---|---|---|---|---|
| **Mode** | Volume (thousands) | Proportion (%) | Volume (thousands) | Proportion (%) | Volume (thousands) | Proportion (%) |
| Transit | 2057.87 | 70.96 | 1837.89 | 67.80 | -10.69 | -4.44 |
| *Subway* | 1494.26 | 51.54 | — | — | — | — |
| *Bus* | 215.78 | 7.42 | — | — | — | — |
| *Railroad* | 297.57 | 10.24 | — | — | — | — |
| *Ferry/Tram* | 51.26 | 1.76 | — | — | — | — |
| Auto (All) | 800.40 | 27.60 | 826.71 | 30.50 | +3.29 | +10.51 |
| *Driving* | — | — | 256.91 | 9.48 | — | — |
| *Carpool* | — | — | 358.72 | 13.23 | — | — |
| *On-demand* | — | — | 211.08 | 7.79 | — | — |
| Non-motorized | 42.01 | 1.45 | 46.11 | 1.70 | +9.77 | +17.44 |
| *Biking* | 42.01 | 1.45 | 21.77 | 0.80 | -48.18 | -44.56 |

[1] https://www.replicahq.com/data-validations. Retrieved January 21, 2026.

[2] https://www.nymtc.org/Data-and-Modeling/Transportation-Data-and-Statistics/Publications/Hub-Bound-Travel. Retrieved January 21, 2026.

| *Walking* | — | — | 24.34 | 0.90 | — | — |
|---|---|---|---|---|---|---|
| **Total** | **2900.28** | **100.00** | **2710.72** | **100.0** | **-6.54** | **+0.00** |

Note: Percent Difference = (Replica − NYMTC) / NYMTC × 100. NYMTC data is used as base. Indented rows show subcategories. "—" indicates data not available or not comparable. In NYMTC data, the category "Auto (All)" includes auto, taxi, van, and truck.

**Table A2**

Comparison of entry sector distribution: NYMTC vs. Replica data

| | **NYMTC Data** | | **Replica Data** | | **Percent Difference** | |
|---|---|---|---|---|---|---|
| **Entry Sector** | Volume (thousands) | Proportion (%) | Volume (thousands) | Proportion (%) | Volume (thousands) | Proportion (%) |
| $60^{th}$ Street | 1071.24 | 36.94 | 982.47 | 36.24 | -8.29 | -1.87 |
| Brooklyn | 761.86 | 26.27 | 725.77 | 26.77 | -4.74 | +1.93 |
| Queens | 570.03 | 19.65 | 581.44 | 21.45 | +2.00 | +9.13 |
| New Jersey | 471.56 | 16.26 | 421.04 | 15.53 | -10.71 | -4.47 |
| Staten Island | 25.61 | 0.88 | — | — | — | — |
| **Total** | **2900.30** | **100.00** | **2710.72** | **100.00** | **-6.54** | **+0.00** |

Note: Percent Difference = (Replica − NYMTC) / NYMTC × 100. NYMTC data is used as base. "—" indicates data not available. The Staten Island Sector enters the CRZ via the Staten Island Ferry service and thus cannot be identified using Replica data.

### A2. Transit travel time retrieved from OpenTripPlanner

Table A3 lists the average transit wait aggregated by trips starting from various regions obtained using OpenTripPlanner.

**Table A3**

Transit times (minutes) and number of transfers aggregated by trip origins

| Origin Region | Transit Waiting Time (min) | | | Transit In-Vehicle Time (min) | | |
|---|---|---|---|---|---|---|
| | 2023 | 2025 | % Change | 2023 | 2025 | % Change |
| CRZ | 1.50 | 1.46 | -2.94% | 8.73 | 8.52 | -2.35% |
| Upper Manhattan | 2.16 | 2.12 | -1.91% | 14.61 | 13.87 | -5.07% |
| Rest of NYC | 4.56 | 4.52 | -0.98% | 23.51 | 23.13 | -1.61% |
| CSA–New York State | 17.06 | 15.04 | -11.82% | 63.67 | 62.51 | -1.82% |
| CSA–New Jersey | 7.00 | 7.20 | +2.80% | 38.77 | 37.54 | -3.17% |

| CSA–Pennsylvania & Connecticut | 16.62 | 16.15 | -2.83% | 83.96 | 83.26 | +0.95% |
|---|---|---|---|---|---|---|

Note: CRZ = Congestion Relief Zone; CSA = Combined Statistical Area. All values represent average travel times in minutes.

## A3. Notations in the choice models and welfare analysis

**Table A4**
Notations used in the choice models and welfare analysis

| | |
|---|---|
| $n \in N$ | Individual index and set of individuals |
| $j \in J$ | Alternative (mode–destination pair) index and choice set |
| $t \in T$ | Market (origin county) index and set of markets |
| $g \in G$ | Trip segment index and set of 16 segments (population × purpose × period) |
| $m \in M$ | Travel mode index and set of modes |
| $d \in D$ | Destination index and set of destinations |
| $h \in H$ | Hierarchical nesting dimension index and set of dimensions |
| $i \in I$ | Population group index; I = {NotLowIncome, LowIncome, Senior, Student} |
| $U_{jt}$ | The total utility trips in market t choosing alternative j |
| $\bar{\delta}_{jt}$ | The mean (systematic) utility of alternative j in market t |
| $V_{jt}^{g}$ | The systematic utility of alternative j in market t for segment g |
| $\varepsilon_{jt}^{g}$ | The random disturbance following Gumbel distribution across j and t |
| $\mu_{njt}$ | The individual-level heterogeneity utility term |
| $X_{jt}^{g}$ | The vector of attributes of alternative j in market t for segment g |
| $s_{jt}^{g}$ | The market share of alternative j in market t for segment g |
| $s_{0t}^{g}$ | The market share of the outside alternative in market t for segment g |
| $o_{t}^{g}$ | The market-specific constant for the outside alternative |
| $V_{m,d,t}^{g}$ | The systematic utility for mode m from market t to destination d |
| $IsNYC_{t}$ | A binary variable equal to 1 if market t is located in New York City |
| $TT_{m,d,t}^{g}$ | The travel time (min) of taking mode m from market t to destination d |
| $CO_{m,d,t}^{g}$ | The monetary cost ($) of taking mode m from market t to destination d |
| $AT_{transit,d,t}^{g}$ | The access time (min) for taking public transit |
| $ET_{transit,d,t}^{g}$ | The egress time (min) for taking public transit |
| $WT_{transit,d,t}^{g}$ | The wait time (min) for taking public transit |
| $IVT_{transit,d,t}^{g}$ | The in-vehicle time (min) for taking public transit |
| $Trans_{transit,d,t}^{g}$ | The number of transfers for taking public transit |
| $\theta_{g,autoTT}, \theta_{g,autoTT}^{NYC}$ | Auto travel time sensitivity and its NYC interaction for segment g |

$\theta_{g,nonautoTT}, \theta_{g,nonautoTT}^{NYC}$ Non-auto travel time sensitivity and its NYC interaction for segment g

$\theta_{g,cost}, \theta_{g,cost}^{NYC}$ Cost sensitivity and its NYC interaction for segment g

$\theta_{g,AT}, \theta_{g,AT}^{NYC}$ Transit access time sensitivity and its NYC interaction for segment g

$\theta_{g,ET}, \theta_{g,ET}^{NYC}$ Transit egress time sensitivity and its NYC interaction for segment g

$\theta_{g,WT}, \theta_{g,WT}^{NYC}$ Transit wait time sensitivity and its NYC interaction for segment g

$\theta_{g,IVT}, \theta_{g,IVT}^{NYC}$ Transit in-vehicle time sensitivity and its NYC interaction for segment g

$\theta_{g,trans}$ Transit transfer sensitivity for segment g

$\theta_{g,asc}^{m}$ Mode-specific constant for mode m and segment g

$\theta_{g,asc}^{d}$ Destination-specific constant for destination d and segment g

$G_j(s_t; \varphi)$ The invertible generating function of market shares in IPDL

$\varphi$ The vector of nesting parameters in IPDL

$\rho_h$ The nesting parameter for hierarchical dimension h

$c_t$ The market-specific constant in the IPDL inversion

$J_h$ The set of alternatives grouped by nesting dimension h

$V_{jt}^{post,g}$ The systematic utility of alternative j in the post-toll period

$X_{jt}^{ex,g}, X_{jt}^{post,g}$ Alternative attributes in the ex-toll and post-toll periods

$\theta_{g}^{ex}, \theta_{g,md}^{post}$ Taste parameters in the ex-toll and post-toll models

$\Delta^{toll} X_{ij}^{g}$ Changes in attributes directly resulting from the congestion toll

$\Delta^{other} X_{ij}^{g}$ Changes in attributes unrelated to the toll (transit, fuel prices)

$\Delta^{toll} \theta_{g,md}$ Toll-induced preference shift in taste parameters

$\theta_{asc-toll}^{CRZ}$ General toll effect on attractiveness of CRZ as a destination

$\theta_{asc-toll}^{m\prime}$ Toll effect on mode m′ attractiveness for trips entering the CRZ

$VOT_{auto_tt}^{g}$ Value of auto travel time for segment g ($/min)

$CS_t^{g}$ Consumer surplus in market t for segment g

$C^{g}$ Unknown constant in the consumer surplus expression

$CV_t^{g}$ The total compensating variation in market t for segment g ($)

$d_t^{g}$ The total trip demand made by segment g in market t

S A subset of utility components (coalition) in the Shapley decomposition

$\omega_{k,t}^{g}$ Shapley value of component k in market t for segment g

cv(S) Partial compensating variation when only components in S are switched on

$CV_t^{toll,g}$ CV attributable to congestion pricing in market t for segment g

$WTC^{KH}$ Wait time compensation under Kaldor–Hicks efficiency (min)

$FDC^{KH}$ Fare discount compensation under Kaldor–Hicks efficiency ($)

$CV_t^{5\%,g}$ CV from a 5% reduction in transit wait time

$CV_t^{remain,g}$ Remaining uncompensated welfare loss after wait time improvement

$FDC_t^{Pareto,i}$ Fare discount for population group i in market t under Pareto improvement

| | |
|---|---|
| $FDC^{Pareto,i}$ | Maximum fare discount for population group i across all markets |
| $G_i$ | Set of trip segments belonging to population group i |

### A4. Derivation of Eq. (13)

This appendix provides a step-by-step derivation of Eq. (13) from Eqs. (4), (11), and (12). Recall that the index $g$ is dropped for simplicity since a separate model is estimated for each trip segment.

***Step 1: Write Eq. (11) for any inside alternative j and the outside alternative j = 0***

From Eq. (11), the systematic utility of any alternative $j \in J$ in market $t$ is:

$$V_{jt} = \ln G_j(\mathrm{s}_t; \varphi) + c_t \tag{A.1}$$

Applying Eq. (11) to the outside alternative ($j = 0$):

$$V_{0t} = \ln G_0(\mathrm{s}_t; \varphi) + c_t \tag{A.2}$$

***Step 2: Simplify $\ln G_0$ for the outside alternative***

Since the outside alternative is the only product of its type (i.e., it does not share a group with any inside alternative on any dimension), applying Eq. (12) to $j = 0$ gives $J_h(0) = \{0\}$ for all $h$, so that $\sum_{q \in J_h(0)} s_{qt} = s_{0t}$. Substituting into Eq. (12):

$$\ln G_0(\mathrm{s}_t; \varphi) = (1 - \textstyle\sum_{h=1}^{H} \rho_h) \ln(s_{0t}) + \textstyle\sum_{h=1}^{H} \rho_h \ln(s_{0t}) = \ln(s_{0t}) \tag{A.3}$$

where the second equality follows because the coefficients sum to unity: $(1 - \sum_{h=1}^{H} \rho_h) + \sum_{h=1}^{H} \rho_h = 1$.

***Step 3: Solve for the market-specific constant $c_t$***

We define the systematic utility of the outside alternative as a market-specific constant, $V_{0t} = o_t$. Substituting Eq. (A.3) into Eq. (A.2):

$$o_t = \ln(s_{0t}) + c_t \quad \Rightarrow \quad c_t = o_t - \ln(s_{0t}) \tag{A.4}$$

***Step 4: Subtract the outside alternative equation from the inside alternative equation***

Subtracting Eq. (A.2) from Eq. (A.1):

$$V_{jt} - V_{0t} = \ln G_j(\mathrm{s}_t; \varphi) - \ln G_0(\mathrm{s}_t; \varphi) \tag{A.5}$$

Note that the constant $c_t$ cancels on both sides.

***Step 5: Substitute the systematic utility expressions***

Since $V_{jt} = \theta^{\top} \mathrm{X}_{jt}$ and $V_{0t} = o_t$, the left-hand side of Eq. (A.5) becomes:

$$V_{jt} - V_{0t} = \theta^{\top} X_{jt} - o_t \tag{A.6}$$

***Step 6: Expand the right-hand side of Eq. (A.5)***

Substituting Eq. (12) for $\ln G_j$ and Eq. (A.3) for $\ln G_0$:

$$\ln G_j(s_t; \varphi) - \ln G_0(s_t; \varphi) = \left(1 - \sum_{h=1}^{H} \rho_h\right) \ln(s_{jt}) + \sum_{h=1}^{H} \rho_h \ln\left(\sum_{q \in J_h(j)} s_{qt}\right) - \ln(s_{0t}) \tag{A.7}$$

We rearrange the right-hand side by adding and subtracting $\sum_{h=1}^{H} \rho_h \ln(s_{jt})$:

$$= \ln(s_{jt}) - \sum_{h=1}^{H} \rho_h \ln(s_{jt}) + \sum_{h=1}^{H} \rho_h \ln\left(\sum_{q \in J_h(j)} s_{qt}\right) - \ln(s_{0t}) \tag{A.8}$$

$$= \ln(s_{jt}) - \ln(s_{0t}) + \sum_{h=1}^{H} \rho_h \left[\ln\left(\sum_{q \in J_h(j)} s_{qt}\right) - \ln(s_{jt})\right] \tag{A.9}$$

$$= \ln\left(\frac{s_{jt}}{s_{0t}}\right) - \sum_{h=1}^{H} \rho_h \ln\left(\frac{s_{jt}}{\sum_{q \in J_h(j)} s_{qt}}\right) \tag{A.10}$$

where the last step uses the property $\ln(a) - \ln(b) = \ln(a/b)$ and reverses the sign inside the logarithm.

***Step 7: Equate and rearrange to obtain Eq. (13)***

Setting Eq. (A.6) equal to Eq. (A.10):

$$\theta^{\top} X_{jt} - o_t = \ln\left(\frac{s_{jt}}{s_{0t}}\right) - \sum_{h=1}^{H} \rho_h \ln\left(\frac{s_{jt}}{\sum_{q \in J_h(j)} s_{qt}}\right) \tag{A.11}$$

Rearranging by moving the nesting variables to the right-hand side:

$$\ln\left(\frac{s_{jt}}{s_{0t}}\right) = \theta^{\top} X_{jt} + \sum_{h=1}^{H} \rho_h \ln\left(\frac{s_{jt}}{\sum_{q \in J_h(j)} s_{qt}}\right) - o_t, \quad \forall j \in J,\ j \neq 0,\ \forall t \in T \tag{13}$$

## A5. Comparison of model specifications

Table A5 summarizes the parameter estimates from the MNL, NL, and IPDL specifications based on total trips without segmentation. The IPDL model achieves a substantially lower NMAE (10.08%) compared to the NL (32.67%) and MNL (73.77%), confirming that capturing substitution patterns across both mode and destination dimensions yields markedly better predictive accuracy. Notably, the CRZ-specific constant is highly significant in both the MNL and NL but becomes insignificant in the IPDL, indicating that the simpler models confound destination-specific preferences with unaccounted substitution patterns that the IPDL properly absorbs through its dual nesting structure.

**Table A5**

Parameter estimates of MNL, NL, and IPDL (each entry represents the average value, and the number in the parenthesis is the standard error).

| | **MNL** | **NL** | **IPDL** |
| --- | --- | --- | --- |

| **Travel time and cost** | | | |
|---|---|---|---|
| Auto travel time | -0.086*** | -0.081*** | -0.079*** |
| ($\theta_{autoTT}$) | (0.000) | (0.000) | (0.002) |
| Transit access time | -0.074*** | -0.112*** | -0.093*** |
| ($\theta_{AT}$) | (0.003) | (0.005) | (0.025) |
| Transit egress time | -0.003 | -0.100*** | -0.103*** |
| ($\theta_{ET}$) | (0.002) | (0.008) | (0.007) |
| Transit wait time | -0.080*** | -0.057*** | -0.683* |
| ($\theta_{WT}$) | (0.010) | (0.002) | (0.324) |
| Transit in-vehicle time | -0.028*** | -0.032*** | -0.041*** |
| ($\theta_{IVT}$) | (0.002) | (0.001) | (0.008) |
| Number of transfers | -0.916*** | -1.318*** | -0.631*** |
| ($\theta_{trans}$) | (0.072) | (0.069) | (0.023) |
| Non-vehicle travel time ($\theta_{nonautoTT}$) | -0.028*** | -0.057*** | -0.018** |
| | (0.000) | (0.002) | (0.006) |
| Trip cost | -0.168*** | -0.239*** | -0.243*** |
| ($\theta_{cost}$) | (0.006) | (0.008) | (0.007) |
| Auto travel time | 0.001** | -0.018*** | -0.027*** |
| ($\theta_{autoTT}^{NYC}$) | (0.000) | (0.001) | (0.001) |
| Transit access time | -0.033*** | -0.021*** | -0.044*** |
| ($\theta_{AT}^{NYC}$) | (0.001) | (0.003) | (0.005) |
| Transit egress time | 0.008*** | -0.061*** | -0.054*** |
| ($\theta_{ET}^{NYC}$) | (0.002) | (0.010) | (0.005) |
| Transit wait time | 0.033*** | -0.033*** | -0.667*** |
| ($\theta_{WT}^{NYC}$) | (0.002) | (0.003) | (0.080) |
| Transit in-vehicle time | 0.001* | -0.016*** | -0.029*** |
| ($\theta_{IVT}^{NYC}$) | (0.001) | (0.002) | (0.002) |
| Non-vehicle travel time ($\theta_{nonautoTT}^{NYC}$) | -0.013*** | -0.011*** | -0.024*** |
| | (0.000) | (0.002) | (0.001) |
| Trip cost | 0.044*** | 0.014*** | 0.029*** |
| ($\theta_{cost}^{NYC}$) | (0.002) | (0.002) | (0.001) |
| **Mode and destination constant** | | | |
| Driving constant | 0.459*** | 0.555*** | 0.491*** |
| ($\theta_{asc}^{driving}$) | (0.048) | (0.049) | (0.045) |
| Transit constant | -0.763** | 0.081 | 0.607* |
| ($\theta_{asc}^{transit}$) | (0.253) | (0.180) | (0.258) |
| FHV constant | -0.098 | 0.819* | 0.055 |
| ($\theta_{asc}^{fhv}$) | (0.274) | (0.370) | (0.152) |
| Biking constant | -1.647*** | -1.553*** | -1.244*** |
| ($\theta_{asc}^{biking}$) | (0.240) | (0198) | (0.183) |
| Walking constant | -0.622*** | -0.126 | -0.127 |
| ($\theta_{asc}^{walking}$) | (0.103) | (0.097) | (0.089) |
| CRZ-specific constant | 0.509*** | -2.848*** | -0.036 |
| ($\theta_{asc}^{CRZ}$) | (0.134) | (0.135) | (0.044) |
| **Nesting parameter** | | | |
| $\ln\left(\frac{s_{jt}}{\sum_{q \in J_{mode}} s_{qt}}\right)$ | — | — | 0.310*** |
| | | | (0.001) |
| $\ln\left(\frac{s_{jt}}{\sum_{q \in J_{destination}} s_{qt}}\right)$ | — | 0.907***<br>(0.009) | 0.234***<br>(0.001) |
| **Meta information** | | | |
| Instrumental variables | Yes | Yes | Yes |
| Market fixed effects | Yes | Yes | Yes |
| # Observations | 46,378 | 46,378 | 46,378 |
| Total trips per day | 70,864,699 | 70,864,699 | 70,864,699 |
| MAE (num. trips) | 1,127.25 | 499.26 | 165.07 |
| NMAE (%) | 73.77% | 32.67% | 10.08% |

Note: ***p-value<0.001, **p-value<0.01, *p-value<0.05. In the NL specification, destination choice is modeled at the upper branch and mode choice at the lower branch. MAE denotes the mean absolute error of alternative-level trip volume predictions, and NMAE denotes the normalized MAE, obtained by scaling MAE by the mean trip volume.

Table A6 compares four IPDL specifications to assess the contribution of market-level fixed effects, NYC-specific interaction terms, and instrumental variables. Removing market fixed effects increases the NMAE from 10.08% to 16.64%, while removing interaction terms raises it to 18.62%, demonstrating that both components contribute meaningfully to model performance. In contrast, the IPDL without instrumental variables achieves a slightly lower NMAE (9.47%) than the full IPDL (10.08%), indicating minimal prediction loss from the endogeneity correction. This small tradeoff is expected, as instrumental variable estimation prioritizes consistent parameter recovery over in-sample fit, and the close agreement between the two specifications supports the validity of the instruments used.

**Table A6**

Parameter estimates of IPDL and IPDL without fixed effects, interaction terms, instrumental variables (each entry represents the average value, and the number in the parenthesis is the standard error).

| | **IPDL (no fixed effect)** | **IPDL (no interaction)** | **IPDL (no IV)** | **IPDL** |
|---|---|---|---|---|
| **Travel time and cost** | | | | |
| Auto travel time | -0.079*** | -0.165*** | -0.089*** | -0.079*** |
| ($\theta_{autoTT}$) | (0.002) | (0.002) | (0.002) | (0.002) |
| Transit access time | -0.093*** | -0.147*** | -0.099*** | -0.093*** |
| ($\theta_{AT}$) | (0.025) | (0.014) | (0.025) | (0.025) |
| Transit egress time | -0.103*** | -0.221*** | -0.104*** | -0.103*** |
| ($\theta_{ET}$) | (0.007) | (0.005) | (0.007) | (0.007) |
| Transit wait time | -0.683* | -0.173 | -0.663* | -0.683* |
| ($\theta_{WT}$) | (0.324) | (0.195) | (0.322) | (0.324) |
| Transit in-vehicle time | -0.041*** | -0.076*** | -0.042*** | -0.041*** |
| ($\theta_{IVT}$) | (0.008) | (0.006) | (0.008) | (0.008) |
| Number of transfers | -0.631*** | -0.111*** | -0.631*** | -0.631*** |
| ($\theta_{trans}$) | (0.023) | (0.022) | (0.023) | (0.023) |
| Non-vehicle travel time | -0.018** | -0.029*** | -0.022*** | -0.018** |
| ($\theta_{nonautoTT}$) | (0.006) | (0.006) | (0.006) | (0.006) |
| Trip cost | -0.243*** | -0.139*** | -0.159*** | -0.243*** |
| ($\theta_{cost}$) | (0.007) | (0.005) | (0.004) | (0.007) |

| **NYC-specific interaction terms** | | | | |
|---|---|---|---|---|
| Auto travel time | -0.027*** | — | -0.027*** | -0.027*** |
| ($\theta_{autoTT}^{NYC}$) | (0.001) | | (0.001) | (0.001) |
| Transit access time | -0.044*** | — | -0.043*** | -0.044*** |
| ($\theta_{AT}^{NYC}$) | (0.005) | | (0.005) | (0.005) |
| Transit egress time | -0.054*** | — | -0.052*** | -0.054*** |
| ($\theta_{ET}^{NYC}$) | (0.005) | | (0.005) | (0.005) |
| Transit wait time | -0.667*** | — | -0.659*** | -0.667*** |
| ($\theta_{WT}^{NYC}$) | (0.080) | | (0.079) | (0.080) |
| Transit in-vehicle time | -0.029*** | — | -0.029*** | -0.029*** |
| ($\theta_{IVT}^{NYC}$) | (0.002) | | (0.002) | (0.002) |
| Non-vehicle travel time | -0.024*** | — | -0.023*** | -0.024*** |
| ($\theta_{nonautoTT}^{NYC}$) | (0.001) | | (0.001) | (0.001) |
| Trip cost | 0.029*** | — | 0.020*** | 0.029*** |
| ($\theta_{cost}^{NYC}$) | (0.001) | | (0.001) | (0.001) |
| **Mode and destination constant** | | | | |
| Driving constant | 0.176*** | 0.127*** | 0.552*** | 0.491*** |
| ($\theta_{asc}^{driving}$) | (0.003) | (0.004) | (0.049) | (0.045) |
| Transit constant | 0.390*** | 0.875*** | 0.427* | 0.607* |
| ($\theta_{asc}^{transit}$) | (0.020) | (0.020) | (0.204) | (0.258) |
| FHV constant | -0.244*** | -0.205*** | 0.295 | 0.055 |
| ($\theta_{asc}^{fhv}$) | (0.021) | (0.018) | (0.194) | (0.152) |
| Biking constant | -1.219*** | -0.975*** | -1.429*** | -1.244*** |
| ($\theta_{asc}^{biking}$) | (0.016) | (0.017) | (0.193) | (0.183) |
| Walking constant | -0.417*** | -0.422*** | 0.023 | -0.127 |
| ($\theta_{asc}^{walking}$) | (0.007) | (0.008) | (0.090) | (0.089) |
| CRZ-specific constant | -0.036 | 0.030 | -0.066 | -0.036 |
| ($\theta_{asc}^{CRZ}$) | (0.044) | (0.048) | (0.044) | (0.044) |
| **Nesting parameter** | | | | |
| $\ln\left(\frac{s_{jt}}{\sum_{q\in J_{mode}} s_{qt}}\right)$ | 0.310*** | 0.293*** | 0.309*** | 0.310*** |
| | (0.001) | (0.001) | (0.001) | (0.001) |
| $\ln\left(\frac{s_{jt}}{\sum_{q\in J_{destination}} s_{qt}}\right)$ | 0.234*** | 0.260*** | 0.235*** | 0.234*** |
| | (0.001) | (0.001) | (0.001) | (0.001) |
| **Meta information** | | | | |
| Instrumental variables | Yes | Yes | No | Yes |
| Market fixed effects | No | Yes | Yes | Yes |
| # Observations | 46,378 | 46,378 | 46,378 | 46,378 |
| Total trips per day | 70,864,699 | 70,864,699 | 70,864,699 | 70,864,699 |
| MAE (num. trips) | 254.22 | 284.57 | 155.08 | 165.07 |
| NMAE (%) | 16.64% | 18.62% | 9.47% | 10.08% |

Note: ***p-value<0.001, **p-value<0.01, *p-value<0.05. MAE denotes the mean absolute error of alternative-level trip volume predictions, and NMAE denotes the normalized MAE, obtained by scaling MAE by the mean trip volume.

As shown in Table A7, The diagnostic tests for the IV regression provide strong evidence supporting the validity and relevance of our instrument specification. The weak instruments tests yield highly significant results (p < 2e-16), indicating that our instruments are strongly correlated with the endogenous cost variables. The Wu-Hausman test statistic of 194.8 (p < 2e-16) strongly rejects the null hypothesis that the cost variables are exogenous, confirming the necessity of using instrumental variables to obtain consistent parameter estimates. The Sargan test for overidentifying restrictions yields a statistic of 482.3 (p < 2e-16). While the significant Sargan test suggests potential concerns about instrument validity, this is common in large samples where even minor specification issues can lead to rejection. Moreover, the goodness-of-fit of the IPDL model with instrumental variables is only slightly lower than that without IV correction, suggesting that addressing the endogeneity of cost variables provides more consistent estimates without sacrificing too much prediction accuracy. Given these evidences, we proceed with the IV specification.

**Table A7**

IV regression diagnostic tests for IPDL

| | **Degree of freedom** | **Statistic** | **p-value** | **Sig.** |
|---|---|---|---|---|
| Weak instruments | 45828 | 10546.1 | <2e-16 | *** |
| Wu-Hausman | 45828 | 194.8 | <2e-16 | *** |
| Sargan | 45828 | 482.3 | <2e-16 | *** |

## A6. Model results for other trip segments

Table A8 summarizes the parameter estimates from the IPDL models for trips made by not-low-income population.

**Table A8**

Parameter estimates of IPDL models for not-low-income population (each entry represents the average value, and the number in the parenthesis is the standard error).

| | **NotLowIncome, Commute, Overnight** | **NotLowIncome, Non-commute, Peak** | **NotLowIncome, Non-commute, Overnight** |
|---|---|---|---|
| **Travel time and cost** | | | |
| Auto travel time | -0.082*** | -0.053*** | -0.032*** |
| ($\theta_{autoTT}$) | (0.006) | (0.008) | (0.006) |
| Transit access time | -0.098*** | -0.099*** | -0.037*** |
| ($\theta_{AT}$) | (0.008) | (0.028) | (0.006) |
| Transit egress time | 0.016 | -0.060*** | -0.055 |
| ($\theta_{ET}$) | (0.012) | (0.013) | (0.033) |
| Transit wait time | -0.414 | -0.821 | -0.446 |
| ($\theta_{WT}$) | (0.369) | (0.526) | (0.560) |
| Transit in-vehicle time | -0.032*** | -0.046 | -0.027 |
| ($\theta_{IVT}$) | (0.005) | (0.121) | (0.034) |
| Number of transfers | -0.574*** | -0.360** | -0.671*** |
| ($\theta_{trans}$) | (0.042) | (0.130) | (0.102) |
| Non-vehicle travel time | -0.039*** | -0.038 | -0.020*** |
| ($\theta_{nonautoTT}$) | (0.005) | (0.025) | (0.004) |
| Trip cost | -0.358*** | -0.207*** | -0.317*** |
| ($\theta_{cost}$) | (0.029) | (0.012) | (0.018) |
| **NYC-specific interaction terms** | | | |
| Auto travel time | -0.024*** | -0.034*** | -0.031*** |
| ($\theta_{autoTT}^{NYC}$) | (0.001) | (0.002) | (0.002) |
| Transit access time | -0.022*** | -0.012*** | -0.009*** |
| ($\theta_{AT}^{NYC}$) | (0.002) | (0.003) | (0.001) |
| Transit egress time | -0.026* | -0.003 | -0.037 |
| ($\theta_{ET}^{NYC}$) | (0.011) | (0.029) | (0.028) |
| Transit wait time | 0.464*** | -0.076** | -0.188 |
| ($\theta_{WT}^{NYC}$) | (0.110) | (0.029) | (0.287) |
| Transit in-vehicle time | -0.021*** | -0.006 | -0.024** |
| ($\theta_{IVT}^{NYC}$) | (0.003) | (0.013) | (0.008) |
| Non-vehicle travel time | -0.005*** | -0.007*** | -0.031*** |
| ($\theta_{nonautoTT}^{NYC}$) | (0.000) | (0.000) | (0.003) |
| Trip cost | 0.071*** | 0.032*** | 0.027*** |
| ($\theta_{cost}^{NYC}$) | (0.005) | (0.002) | (0.001) |
| **Mode and destination constant** | | | |
| Driving constant | 0.323*** | 0.383*** | 0.370*** |
| ($\theta_{asc}^{driving}$) | (0.012) | (0.011) | (0.012) |
| Transit constant | 0.500*** | 0.245* | 0.024 |
| ($\theta_{asc}^{transit}$) | (0.046) | (0.100) | (0.085) |
| FHV constant | -0.963*** | -0.548*** | -0.563*** |
| ($\theta_{asc}^{fhv}$) | (0.058) | (0.039) | (0.037) |
| Biking constant | -2.000*** | -1.756*** | -1.604*** |
| ($\theta_{asc}^{biking}$) | (0.066) | (0.056) | (0.055) |
| Walking constant | -0.715*** | -0.323*** | -0.361*** |
| ($\theta_{asc}^{walking}$) | (0.025) | (0.023) | (0.020) |
| CRZ-specific constant | -0.387*** | 0.017 | -0.155 |
| ($\theta_{asc}^{CRZ}$) | (0.101) | (0.134) | (0.112) |
| **Nesting parameter** | | | |
| $\ln\left(\frac{s_{jt}}{\sum_{q\in J_{mode}} s_{qt}}\right)$ | 0.293*** | 0.321*** | 0.321*** |
| | (0.002) | (0.002) | (0.002) |
| $\ln\left(\frac{s_{jt}}{\sum_{q\in J_{destination}} s_{qt}}\right)$ | 0.159*** | 0.181*** | 0.171*** |
| | (0.003) | (0.003) | (0.003) |
| **Meta information** | | | |
| Instrumental variables | Yes | Yes | Yes |
| Market fixed effect | Yes | Yes | Yes |
| # Observations | 3,043 | 3,218 | 3,083 |
| # Total trips per day | 1,264,670 | 18,199,620 | 3,386,751 |
| MAE (num. trips) | 21.43 | 218.01 | 50.59 |
| NMAE (%) | 5.16% | 3.85% | 4.61% |

Note: ***p-value<0.001, **p-value<0.01, *p-value<0.05. Given the table length, only the destination constant of CRZ is reported in the table. MAE denotes the mean absolute error of alternative-level trip volume predictions, and NMAE denotes the normalized MAE, obtained by scaling MAE by the mean trip volume.

Table A9 summarizes the parameter estimates from the IPDL models for trips made by low-income population.

**Table A9**
Parameter estimates of IPDL models for low-income population (each entry represents the average value, and the number in the parenthesis is the standard error).

| | LowIncome, Commute, Overnight | LowIncome, Non-commute, Peak | LowIncome, Non-commute, Overnight |
|---|---|---|---|
| **Travel time and cost** | | | |
| Auto travel time ($\theta_{autoTT}$) | -0.069*** (0.006) | -0.054*** (0.009) | -0.040*** (0.007) |
| Transit access time ($\theta_{AT}$) | -0.122*** (0.009) | -0.148*** (0.027) | -0.022*** (0.003) |
| Transit egress time ($\theta_{ET}$) | -0.025* (0.013) | -0.116*** (0.032) | -0.062* (0.030) |
| Transit wait time ($\theta_{WT}$) | -0.518 (0.360) | -1.419 (2.060) | 0.291 (0.835) |
| Transit in-vehicle time ($\theta_{IVT}$) | -0.041*** (0.005) | 0.025 (0.023) | -0.014 (0.013) |
| Number of transfers ($\theta_{trans}$) | -0.316*** (0.044) | -0.684*** (0.105) | -0.678*** (0.087) |
| Non-vehicle travel time ($\theta_{nonautoTT}$) | -0.036*** (0.005) | -0.035 (0.019) | -0.041*** (0.007) |
| Trip cost ($\theta_{cost}$) | -0.547*** (0.080) | -0.667*** (0.071) | -0.535*** (0.057) |
| **NYC-specific interaction terms** | | | |
| Auto travel time ($\theta_{autoTT}^{NYC}$) | -0.030*** (0.002) | -0.034*** (0.002) | -0.032*** (0.002) |
| Transit access time ($\theta_{AT}^{NYC}$) | -0.013*** (0.001) | -0.020*** (0.003) | -0.019*** (0.002) |
| Transit egress time ($\theta_{ET}^{NYC}$) | -0.031** (0.012) | -0.031 (0.025) | -0.022 (0.026) |
| Transit wait time ($\theta_{WT}^{NYC}$) | 0.716*** (0.112) | 0.005 (0.452) | 0.636** (0.222) |
| Transit in-vehicle time ($\theta_{IVT}^{NYC}$) | -0.023*** (0.003) | -0.026** (0.010) | -0.027*** (0.006) |
| Non-vehicle travel time ($\theta_{nonautoTT}^{NYC}$) | -0.003*** (0.000) | -0.032*** (0.003) | -0.019*** (0.002) |
| Trip cost ($\theta_{cost}^{NYC}$) | 0.065*** (0.008) | 0.093*** (0.009) | 0.084*** (0.007) |
| **Mode and destination constant** | | | |
| Driving constant ($\theta_{asc}^{driving}$) | 0.197*** (0.013) | 0.286*** (0.011) | 0.275*** (0.012) |
| Transit constant ($\theta_{asc}^{transit}$) | 0.744*** (0.049) | 0.190* (0.083) | 0.195** (0.075) |
| FHV constant ($\theta_{asc}^{fhv}$) | -1.430*** (0.075) | -0.953*** (0.058) | -0.966*** (0.056) |
| Biking constant ($\theta_{asc}^{biking}$) | -1.729*** (0.061) | -1.388*** (0.047) | -1.241*** (0.047) |
| Walking constant | -0.375*** | -0.168*** | -0.128*** |
| ($\theta_{asc}^{walking}$) | (0.026) | (0.022) | (0.021) |
| CRZ-specific constant | -0.466*** | -0.061 | -0.329** |
| ($\theta_{asc}^{CRZ}$) | (0.113) | (0.139) | (0.119) |
| **Nesting parameter** | | | |
| $\ln\left(\frac{s_{jt}}{\sum_{q\in J_{mode}} s_{qt}}\right)$ | 0.297*** | 0.321*** | 0.318*** |
| | (0.002) | (0.002) | (0.002) |
| $\ln\left(\frac{s_{jt}}{\sum_{q\in J_{destination}} s_{qt}}\right)$ | 0.149*** | 0.177*** | 0.164*** |
| | (0.003) | (0.003) | (0.003) |
| **Meta information** | | | |
| Instrumental variables | Yes | Yes | Yes |
| Market fixed effect | Yes | Yes | Yes |
| # Observations | 2,690 | 3,034 | 2,860 |
| # Total trips per day | 427,302 | 6,146,422 | 1,162,135 |
| MAE (num. trips) | 8.87 | 87.43 | 21.68 |
| NMAE (%) | 5.59% | 4.32% | 5.34% |

Note: ***p-value<0.001, **p-value<0.01, *p-value<0.05. Given the table length, only the destination constant of CRZ is reported in the table. MAE denotes the mean absolute error of alternative-level trip volume predictions, and NMAE denotes the normalized MAE, obtained by scaling MAE by the mean trip volume.

Table A10 summarizes the parameter estimates from the IPDL models for trips made by senior population.

**Table A10**

Parameter estimates of IPDL models for senior population (each entry represents the average value, and the number in the parenthesis is the standard error).

| | **Senior, Commute, Peak** | **Senior, Commute, Overnight** | **Senior, Non-commute, Overnight** |
|---|---|---|---|
| **Travel time and cost** | | | |
| Auto travel time | -0.068*** | -0.035*** | -0.020*** |
| ($\theta_{autoTT}$) | (0.006) | (0.004) | (0.005) |
| Transit access time | -0.206*** | -0.103*** | -0.052** |
| ($\theta_{AT}$) | (0.023) | (0.010) | (0.018) |
| Transit egress time | -0.218*** | -0.043* | -0.095** |
| ($\theta_{ET}$) | (0.017) | (0.018) | (0.035) |
| Transit wait time | -1.086 | 0.314 | -0.322 |
| ($\theta_{WT}$) | (0.892) | (0.635) | (0.984) |
| Transit in-vehicle time | -0.046*** | -0.023** | 0.039 |
| ($\theta_{IVT}$) | (0.011) | (0.007) | (0.051) |
| Number of transfers | -0.046 | -0.408*** | -0.721*** |
| ($\theta_{trans}$) | (0.058) | (0.048) | (0.114) |
| Non-vehicle travel time | -0.041*** | -0.031*** | -0.052*** |
| ($\theta_{nonautoTT}$) | (0.006) | (0.006) | (0.010) |
| Trip cost | -0.481*** | -0.423*** | -0.396*** |
| ($\theta_{cost}$) | (0.033) | (0.028) | (0.027) |
| **NYC-specific interaction terms** | | | |
| Auto travel time ($\theta_{autoTT}^{NYC}$) | -0.015*** (0.002) | -0.019*** (0.001) | -0.025*** (0.002) |
| Transit access time ($\theta_{AT}^{NYC}$) | -0.011*** (0.001) | -0.014*** (0.001) | -0.008*** (0.002) |
| Transit egress time ($\theta_{ET}^{NYC}$) | 0.034* (0.014) | -0.030 (0.017) | 0.081 (0.065) |
| Transit wait time ($\theta_{WT}^{NYC}$) | -0.043 (0.212) | 0.363* (0.166) | 0.313 (0.464) |
| Transit in-vehicle time ($\theta_{IVT}^{NYC}$) | -0.022*** (0.005) | -0.031*** (0.004) | -0.024* (0.012) |
| Non-vehicle travel time ($\theta_{nonautoTT}^{NYC}$) | -0.006*** (0.000) | -0.005*** (0.000) | -0.008*** (0.000) |
| Trip cost ($\theta_{cost}^{NYC}$) | 0.094*** (0.006) | 0.086*** (0.005) | 0.103*** (0.006) |
| **Mode and destination constant** | | | |
| Driving constant ($\theta_{asc}^{driving}$) | 0.344*** (0.010) | 0.344*** (0.011) | 0.392*** (0.012) |
| Transit constant ($\theta_{asc}^{transit}$) | 0.826*** (0.053) | 0.452*** (0.051) | -0.175 (0.101) |
| FHV constant ($\theta_{asc}^{fhv}$) | -0.992*** (0.049) | -0.784*** (0.041) | -0.687*** (0.040) |
| Biking constant ($\theta_{asc}^{biking}$) | -2.202*** (0.072) | -2.211*** (0.081) | -1.681*** (0.060) |
| Walking constant ($\theta_{asc}^{walking}$) | -0.656*** (0.032) | -0.678*** (0.035) | -0.326*** (0.027) |
| CRZ-specific constant ($\theta_{asc}^{CRZ}$) | 0.045 (0.106) | -0.167 (0.087) | -0.234* (0.105) |
| **Nesting parameter** | | | |
| $\ln\left(\frac{s_{jt}}{\sum_{q\in J_{mode}} s_{qt}}\right)$ | 0.306*** (0.002) | 0.313*** (0.002) | 0.326*** (0.002) |
| $\ln\left(\frac{s_{jt}}{\sum_{q\in J_{destination}} s_{qt}}\right)$ | 0.160*** (0.003) | 0.149*** (0.003) | 0.171*** (0.003) |
| **Meta information** | | | |
| Instrumental variables | Yes | Yes | Yes |
| Market fixed effect | Yes | Yes | Yes |
| # Observations | 3,022 | 2,706 | 2,710 |
| # Total trips per day | 2,457,390 | 313,212 | 827,002 |
| MAE (num. trips) | 33.64 | 5.96 | 13.48 |
| NMAE (%) | 4.14% | 5.15% | 4.42% |

Note: ***p-value<0.001, **p-value<0.01, *p-value<0.05. Given the table length, only the destination constant of CRZ is reported in the table. MAE denotes the mean absolute error of alternative-level trip volume predictions, and NMAE denotes the normalized MAE, obtained by scaling MAE by the mean trip volume.

Table A11 summarizes the parameter estimates from the IPDL models for trips made by student population.

**Table A11**
Parameter estimates of IPDL models for student population (each entry represents the average value, and the number in the parenthesis is the standard error).

| | **Student, Commute, Peak** | **Student, Commute, Overnight** | **Student, Non-commute, Overnight** |
|---|---|---|---|
| **Travel time and cost** | | | |
| Auto travel time ($\theta_{autoTT}$) | -0.074*** (0.009) | -0.050*** (0.006) | -0.027*** (0.006) |
| Transit access time ($\theta_{AT}$) | -0.230*** (0.039) | -0.163*** (0.020) | -0.029*** (0.005) |
| Transit egress time ($\theta_{ET}$) | -0.243*** (0.028) | -0.020 (0.017) | -0.026 (0.015) |
| Transit wait time ($\theta_{WT}$) | -1.612 (1.709) | 0.084 (0.547) | -0.240 (0.323) |
| Transit in-vehicle time ($\theta_{IVT}$) | -0.047* (0.019) | -0.040*** (0.008) | -0.023 (0.038) |
| Number of transfers ($\theta_{trans}$) | -0.471*** (0.109) | -0.404*** (0.052) | -0.714*** (0.097) |
| Non-vehicle travel time ($\theta_{nonautoTT}$) | -0.054*** (0.004) | -0.027*** (0.004) | -0.028*** (0.006) |
| Trip cost ($\theta_{cost}$) | -0.778*** (0.078) | -0.533*** (0.061) | -0.381*** (0.031) |
| **NYC-specific interaction terms** | | | |
| Auto travel time ($\theta_{autoTT}^{NYC}$) | -0.028*** (0.003) | -0.033*** (0.002) | -0.030*** (0.002) |
| Transit access time ($\theta_{AT}^{NYC}$) | -0.016*** (0.002) | -0.012*** (0.001) | -0.025*** (0.003) |
| Transit egress time ($\theta_{ET}^{NYC}$) | -0.077*** (0.023) | -0.044** (0.017) | -0.074 (0.042) |
| Transit wait time ($\theta_{WT}^{NYC}$) | -0.029 (0.402) | 0.259 (0.162) | 0.363 (0.330) |
| Transit in-vehicle time ($\theta_{IVT}^{NYC}$) | -0.037*** (0.009) | -0.028*** (0.004) | -0.034*** (0.009) |
| Non-vehicle travel time ($\theta_{nonautoTT}^{NYC}$) | -0.005*** (0.000) | -0.003*** (0.000) | -0.012*** (0.001) |
| Trip cost ($\theta_{cost}^{NYC}$) | 0.118*** (0.014) | 0.078*** (0.008) | 0.101*** (0.007) |
| **Mode and destination constant** | | | |
| Driving constant ($\theta_{asc}^{driving}$) | -0.343*** (0.012) | 0.297*** (0.013) | 0.354*** (0.012) |
| Transit constant ($\theta_{asc}^{transit}$) | 0.477*** (0.085) | 0.496*** (0.052) | 0.082 (0.084) |
| FHV constant ($\theta_{asc}^{fhv}$) | -1.550*** (0.091) | -1.174*** (0.066) | -0.842*** (0.048) |
| Biking constant ($\theta_{asc}^{biking}$) | -1.155*** (0.032) | -1.792*** (0.060) | -1.551*** (0.054) |
| Walking constant ($\theta_{asc}^{walking}$) | -0.085*** (0.023) | -0.361*** (0.025) | -0.253*** (0.022) |
| CRZ-specific constant ($\theta_{asc}^{CRZ}$) | 0.025 (0.166) | -0.214 (0.139) | -0.109 (0.136) |
| **Nesting parameter** | | | |
| $\ln\left(\frac{s_{jt}}{\sum_{q \in J_{mode}} s_{qt}}\right)$ | 0.320*** (0.002) | 0.310*** (0.002) | 0.325*** (0.002) |
| $\ln\left(\frac{s_{jt}}{\sum_{q \in J_{destination}} s_{qt}}\right)$ | 0.187*** (0.003) | 0.156*** (0.004) | 0.166*** (0.003) |
| **Meta information** | | | |
| Instrumental variables | Yes | Yes | Yes |
| Market fixed effect | Yes | Yes | Yes |
| # Observations | 2,923 | 2,385 | 2,590 |
| # Total trips per day | 8,937,077 | 239,864 | 619,289 |
| MAE (num. trips) | 97.27 | 5.98 | 11.14 |
| NMAE (%) | 3.18% | 5.94% | 4.66% |

Note: ***p-value<0.001, **p-value<0.01, *p-value<0.05. Given the table length, only the destination constant of CRZ is reported in the table. MAE denotes the mean absolute error of alternative-level trip volume predictions, and NMAE denotes the normalized MAE, obtained by scaling MAE by the mean trip volume.

### A7. Evaluations under Pareto improvement (55-100% wait time reduction)

Table A12 summarizes the required segment-level fare discounts and corresponding annual subsidies under various levels of wait time reduction from 55-100% for both NYC and New Jersey residents.

**Table A12**
Compensatory transit strategies under Pareto improvement (55-100% wait time reduction)

| Wait time reduction (min) | **NYC part** | | **New Jersey part** | |
|---|---|---|---|---|
| | Segment-level fare discount ($/trip) | Subsidy for discount (M $/year) | Segment-level fare discount ($/trip) | Subsidy for discount (M $/year) |
| 55 | NotLowIncome: 0<br>LowIncome: 0<br>Senior: 0.05<br>Student: 0 | 7.66 | NotLowIncome: 0.37<br>LowIncome: 0.28<br>Senior: 0.70<br>Student: 0.69 | 67.17 |
| 60 | NotLowIncome: 0<br>LowIncome: 0<br>Senior: 0<br>Student: 0 | 0.48 | NotLowIncome: 0.35<br>LowIncome: 0.28<br>Senior: 0.65<br>Student: 0.68 | 63.52 |
| 65 | NotLowIncome: 0<br>LowIncome: 0<br>Senior: 0<br>Student: 0 | 0 | NotLowIncome: 0.33<br>LowIncome: 0.24<br>Senior: 0.60<br>Student: 0.68 | 60.13 |
| 70 | NotLowIncome: 0<br>LowIncome: 0<br>Senior: 0<br>Student: 0 | 0 | NotLowIncome: 0.31<br>LowIncome: 0.24<br>Senior: 0.56<br>Student: 0.67 | 56.74 |
| 75 | NotLowIncome: 0<br>LowIncome: 0<br>Senior: 0<br>Student: 0 | 0 | NotLowIncome: 0.29<br>LowIncome: 0.24<br>Senior: 0.51<br>Student: 0.67 | 53.38 |
| 80 | NotLowIncome: 0<br>LowIncome: 0<br>Senior: 0<br>Student: 0 | 0 | NotLowIncome: 0.27<br>LowIncome: 0.24<br>Senior: 0.48<br>Student: 0.66 | 50.11 |
| 85 | NotLowIncome: 0<br>LowIncome: 0<br>Senior: 0<br>Student: 0 | 0 | NotLowIncome: 0.24<br>LowIncome: 0.24<br>Senior: 0.45<br>Student: 0.65 | 46.90 |
| 90 | NotLowIncome: 0<br>LowIncome: 0<br>Senior: 0<br>Student: 0 | 0 | NotLowIncome: 0.22<br>LowIncome: 0.24<br>Senior: 0.43<br>Student: 0.65 | 43.80 |
| 95 | NotLowIncome: 0<br>LowIncome: 0<br>Senior: 0<br>Student: 0 | 0 | NotLowIncome: 0.20<br>LowIncome: 0.24<br>Senior: 0.42<br>Student: 0.64 | 40.70 |
| 100 | NotLowIncome: 0<br>LowIncome: 0<br>Senior: 0<br>Student: 0 | 0 | NotLowIncome: 0.18<br>LowIncome: 0.20<br>Senior: 0.40<br>Student: 0.64 | 37.60 |